\documentclass[aps,superscriptaddress,nofootinbib]{revtex4}
\def\be{\begin{eqnarray} &&}
\newcommand{\bm}[1] {#1}
\def\nonu{\nonumber \\ &&}
\def\ee{\end{eqnarray}}
 \usepackage{wrapfig}
 \usepackage[normalem]{ulem}
 \usepackage{graphicx,color}
 \usepackage{enumerate}

     \font\tenbifull=cmmib10 scaled 1200 
     \font\tenbimed=cmmib9
     \font\tenbismall=cmmib7
       \def\bmit{\fam9 }

\mathchardef\bbkappa="7114
\mathchardef\bbrho="711A
\mathchardef\bbsigma="711B
\mathchardef\bbtau="711C
\mathchardef\bbvarrho="7125
\mathchardef\bbvarsigma="7126
\mathchardef\bbxi="7118

\def\boldsigma{{\bmit\bbsigma}}

\newcommand\la{\langle}
\newcommand\ra{\rangle}

\makeindex

\def \bfgr #1{ \mbox {{\boldmath $#1$}}}
\def\beq{\begin{equation}}
\def\eeq{\end{equation}}
\usepackage{amsbsy,amsmath}
\newcommand{\bs}[1]{\boldsymbol{#1}}
\newcommand{\KE}{\bs{\mathcal{K}}}

\begin{document}


 \date{today}
\vskip 2mm
\date{\today}
\vskip 2mm
\title{Distorted spin-dependent spectral function
of an $A=3$ nucleus and semi-inclusive deep inelastic scattering processes}

 \author{L.P. Kaptari}
 \affiliation{Department of Physics, University of Perugia  and\\
    Istituto Nazionale di Fisica Nucleare, Sezione di Perugia,
    Via A. Pascoli, I-06123, Italy}
 \affiliation{Thomas Jefferson National Accelerator Facility
Newport News, Virginia 23606, USA.}
 \affiliation{Bogoliubov Lab. Theor. Phys., 141980, JINR, Dubna,
 Russia}
\author{A. Del Dotto}
 \affiliation{Dept. of Phys. and Math. Univ. of Rome "Roma Tre" and
    Istituto Nazionale di Fisica Nucleare, Sezione di
    "Roma Tre",
  Via della Vasca Navale 84,  00146 Rome,  Italy}
 \author{E. Pace}
 \affiliation{Phys. Dept. Univ. of Rome "Tor Vergata" and
    Istituto Nazionale di Fisica Nucleare, Sezione di
    Tor Vergata,
  Via della Ricerca Scientifica 1, I-00133, Rome,  Italy}

 \author{G. Salm\`e}
  \affiliation{
   Istituto Nazionale di Fisica Nucleare, Sezione di Roma,
  P-le A. Moro 2, I-00185, Rome,  Italy}

  \author{S. Scopetta}
  \affiliation{Department of Physics, University of Perugia  and\\
     Istituto Nazionale di Fisica Nucleare, Sezione di Perugia,
     Via A. Pascoli, I-06123, Italy}

\begin{abstract}
{  The distorted spin-dependent spectral function
of a nucleon inside an A=3 nucleus is introduced as a novel tool for
investigating
the
polarized electron scattering off polarized $^3$He in semi-inclusive DIS regime
(SiDIS), going beyond the
standard plane wave impulse approximation.
This  distribution function is applied to the study of
the   spectator SiDIS,  $\vec{^3{\rm He}}(\vec e, e' ~{^2}{\rm H})X$,
in order  to properly take into account the final state
interaction between the hadronizing quark and the
detected deuteron, with the final goal  of a more reliable extraction
of the polarized parton-distribution $g_1(x)$ inside a bound { proton}.
Our analysis allows to single out two well-defined  kinematical regions
where the experimental asymmetries could yield very interesting
information:
the region where the final state
effects
can be minimized, and therefore  the { direct} access to the
parton distributions in the   { proton} is feasible, and the  one
 where the final state interaction dominates, and
the spectator SiDIS reactions can elucidate the mechanism of the quark hadronization
itself.
The perspectives of extending our approach
{ i) to the mirror nucleus, $^3$H, for achieving a less model-dependent flavor decomposition,
and ii) to the asymmetries measured in the
standard SiDIS reactions,
$\vec e + \vec{^3 {\rm He}} \to e' + h+X$ with $h$ a detected fast hadron,
with the aim of extracting } the neutron transversity, are discussed.}

\end{abstract}

\date{\today}


\maketitle

\maketitle
\large
\vskip 2cm
\newpage

\section{Introduction}

Inclusive Deep Inelastic Scattering (DIS), i.e. the process $A(l,l')X$,
where {  a lepton interacts with a hadronic target $A$ and only the scattered
lepton is detected},
has been the subject of intense experimental investigation
in the last decades. Through these processes it has become possible
to describe the longitudinal (with respect to the direction of the high energy
beam) momentum and helicity distributions of the partons,
in both proton (see, e.g., Refs.
\cite{unpol, polarized}) and nuclear targets
(see, e.g., Refs. \cite{Arneodo,Piller,emclast}).

Despite of these efforts it is nowadays clear that the answers
to a few crucial questions
can be hardly found through DIS experiments.
Among them, we remind at least three   challenging problems, which are related
to the subject of this paper:
$i)$ the { fully quantitative} explanation of the so-called EMC effect
(i.e. the modification of the nucleon partonic structure due to the nuclear
medium,
observed long time ago \cite{EMC});
$ii)$ the solution of the so-called ``spin crisis'', i.e.
the fact that the nucleon spin
does not originate from
only the spins of  its valence quarks, already reported
in Ref. \cite{polEMC}; $iii)$ the measurement of the  {  chiral-odd  parton
distribution function  (PDF) called  transversity
(see, e.g., Refs. \cite{barone,barone1} { for an introduction to this issue}),
 related to  the    probability   distribution to find a
transversely-polarized quark  inside a transversely-polarized  nucleon, that
allows one to complete the leading-twist description of a
polarized nucleon, unfortunately not measurable
in DIS.}

\begin{figure}[!ht]
\centerline{
\includegraphics[scale=0.35,angle=0]{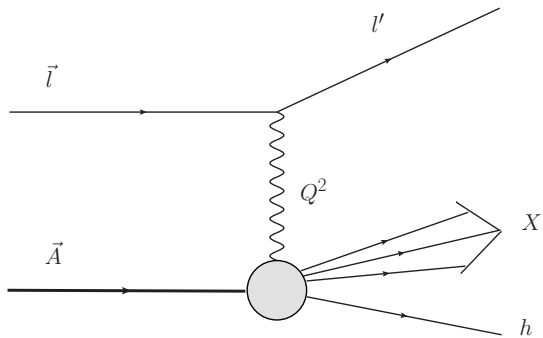}}
\caption{One-photon exchange diagram for
semi-inclusive deep inelastic scattering
processes $\vec A(\vec l,l'\, h) X$. {  The target $A$ can be
either longitudinally or transversely polarized}.}

\label{Fig1}
\end{figure}

  To clarify { the above mentioned  problems,  a new direction has been
  taken. Special
efforts on both experimental and theoretical sides have been focused on
exclusive and semi-inclusive
deep inelastic (SiDIS) measurements, in order
   to go   beyond the
inclusive processes.}
In the  SiDIS reactions, besides the scattered lepton,
a hadron is detected in coincidence, see Fig.\ref{Fig1}.
If the detected hadron is fast,
one can expect that it originates from the
fragmentation of the active, highly   off-mass-shell  quark,
after absorbing
the virtual photon. Hence, the { detected}  hadron
brings information about the motion of quarks
in the parent nucleon before interacting with the photon,    and
in particular on their transverse motion.
 { Therefore, through SiDIS reactions, one can }
 access  the so-called transverse
momentum dependent parton distributions
(TMDs) (see, e.g., Ref. \cite{barone,barone1}) {    establishing  a
framework where one can develop new tools for gathering  a   wealth of 
information on the 
partonic dynamics,
more rich
than in  the collinear
case.
For instance, beside the main issue of TMDs, one should remind that the detected hadron carries { also}
information on the hadronization mechanism, itself.}
{ The SiDIS  cross sections can be parametrized, at leading twist,
by six parton distributions;
this number reduces to three in DIS and increases  to eight
once the so-called time-reversal odd TMDs  are considered
 \cite{barone}.}
{    In view of this},  SiDIS reactions play a  crucial role for
addressing the
issue of the chiral-odd transversity, since { it is   necessary
to flip   the quark
chirality for observing this TMD. Finally, it should be emphasized that
in order to experimentally investigate
the  wide field of  TMDs, one should measure
   cross-section asymmetries,
using different combinations of beam and  target polarizations (see, e.g., Ref. \cite{d'alesio}). Moreover,
 to make complete the study of   TMDs one should achieve
 a sound  flavor decomposition, that  can be obtained once
 neutron data become available.  To this end, it should be pointed out that
 the investigation
of the neutron content is highly favored  by choosing a polarized $^3$He as a
target, as
discussed in detail in what follows.}

As it is well known, free neutron targets are not available
and nuclei have to be used as effective neutron targets.
Due to its peculiar spin structure
(see, e.g., Ref. \cite{friar}),
polarized $^3$He has been used extensively in DIS studies.
In particular, procedures to extract the
neutron spin-dependent structure functions from $^3$He
data, taking properly into account the Fermi-motion and binding effects,
have been proposed and   successfully applied
\cite{antico}. {  Such a detailed description of the target nucleus
was accomplished in plane wave impulse approximation
(PWIA) by using
the so-called spin-dependent spectral function, { whose diagonal elements yield} the probability
distribution to find a nucleon with a given momentum, missing energy and {
polarization}
inside the nucleus. It is worth noting that, within PWIA, accurate
 $^3$He spin-dependent spectral functions,} based on
realistic few-body
calculations (for both the target nucleus and the  spectator
pair in the final state),
have been built in the last twenty years
\cite{cda,SS,cda1,pskv}.   It is very important to notice that
the spin-dependent spectral function contains two  contributions:
one  does not depend upon the polarization of the
target nucleus, ${\bf S}_A$, and the other does.
This is the term that enters  the description of the asymmetries of the cross sections we
are going to investigate, and for the sake of brevity we call it spin-dependent contribution.

The question whether  similar procedures, based on the PWIA spectral function, can be extended to
SiDIS is of great relevance, due to several experiments that exploit
a  polarized $^3$He target  (see, e.g., Ref.   \cite{He3exp}),
for  investigating  the transverse degrees of freedom of the neutron.
For instance,  a wide interest has arisen about the possibility
to measure, using transversely polarized $^3$He,
azimuthal single-spin asymmetries of the neutron,
which are sensitive to i) the so-called
time-reversal odd TMDs (such as the Sivers
\cite{sivers} and Boer-Mulders \cite{boemu} functions) and ii)
fragmentation functions (such as
the Collins function \cite{collins}). { Indeed, it is }
the existence of final state interactions (FSI) at leading twist { that allows} for
time-reversal odd TMDs~\cite{Brodsky:2002cx}.
The first measurements of these functions,
obtained through
SiDIS off transversely polarized
proton and deuteron targets, have produced
a puzzling experimental scenario \cite{hermes,compass}.
Therefore,
with the aim of extracting the neutron information
to shed some light on the problem,
a measurement of SiDIS
off transversely polarized $^3$He has been addressed \cite{bro},
and an experiment, planned to measure azimuthal asymmetries
in the production of leading $\pi^\pm$  from transversely
polarized $^3$He, has been already completed at Jefferson Lab (JLab),
with a beam energy of 6 GeV
\cite{prljlab}. {  Notice that a new  experiment
has been planned and it will be soon performed  after completing the 12
GeV upgrade \cite{jlab12}.}

In view of these experimental { efforts}, a realistic PWIA analysis of SiDIS
off transversely polarized $^3$He has been presented in Ref.
\cite{mio}. A spin-dependent spectral function {  \cite{pskv}, corresponding to the
nucleon-nucleon }
 AV18 interaction \cite{av18}, has been used for a realistic
description of the nuclear dynamics and
the crucial issue of extracting
the neutron information from $^3$He data
has been discussed.
  It was found that { for SiDIS reactions, where both   parton
distributions and
fragmentation functions  are involved}, one can safely extend a
 model-independent { extraction} procedure, based
on the realistic evaluation
of the proton and neutron polarizations in $^3$He
\cite{antico} { and  widely used in inclusive
DIS, given its ability  } to take into account effectively
the momentum and energy distributions
of the polarized bound nucleons in $^3$He.

However, in a SiDIS process the effect of FSI cannot
be neglected ``a priori'' and it needs to be seriously
investigated. This will be the aim of the present paper,
where a distorted spin-dependent spectral function is introduced,
by applying a
generalized eikonal approximation (GEA),
already successfully exploited for describing
unpolarized SiDIS off nuclei \cite{ourlast}.

Concerning the mechanism of SiDIS off nuclear targets,
one can distinguish at least
two, rather different,
  sets of reactions:  \begin{enumerate} [i)]
\item the  most familiar, {\em standard}  process,
where the  fast hadron is  detected mostly in the
forward direction. This implies that the hadron  was produced by the leading
quark. Therefore  such reactions can be used to investigate
TMDs inside the hit nucleon;
\item the {\em spectator}  SiDIS, in which  there is no
 detection of any produced
(fast) hadrons, while
the remaining, slow  $(A-1)$ nucleon system, acting as a spectator of the
photon-nucleon interaction, is detected.\end{enumerate}
It has been shown
that  spectator
 SiDIS can be very
useful to investigate the unpolarized deep inelastic structure
functions $F_{1,2}(x)$ of a bound nucleon, and therefore
to clarify the origin of the EMC effect
\cite{scopetta,simula,smss,ourlast,ckk,veronica}.
At the same time, this process can provide also useful
information on quark hadronization in
medium, complementary to the ones obtained so far by the analysis of the
standard SiDIS process. {  Notably, including polarization degrees of
freedom
 would provide
information on the spin-dependent structure functions $g_{1,2}(x)$
for  bound nucleons and, ultimately,
on the origin of the polarized EMC effect. }

From the theoretical point of view, an important observation is that
in both processes, the standard and spectator SiDISs, nuclear effects
can be described by the same quantity:
a {\em distorted spin-dependent spectral function}, { once FSI are taken into account}.
With respect to
the one governing the DIS reactions,
this { novel distribution function}
is a more complicated object, since,  besides   the momentum, energy and polarization
distributions of the nucleons in $^3$He, it includes
also effects of FSI between the produced particles.
A thorough knowledge of
the distorted spin-dependent spectral function
would allow one to   reliably separate  the effects due to the
nuclear structure from the ones involving TMDs, in the  experimental cross
sections.
Since nuclear effects are expected to be similar in both  SiDIS reactions,
the spectral function and FSI effects   will  be studied  by considering,
as a first step,
the simplest process, i.e., the {\em spectator} SiDIS,
described by { only two   structure functions of the nucleon}.
Then, one can proceed with the more complicated and involved {\em standard} SiDIS,
  focusing  on the extraction of   quark
TMDs inside the neutron, i.e. the needed ingredients for { making complete }
the flavor
decomposition.

  Aim of our paper, is the evaluation of the  distorted spin-dependent
spectral function of  $^3$He, taking into account, through a
generalized  eikonal
approximation (see, e.g., Ref. \cite{ourlast}), the
FSI between the $(A-1)$ spectator and the quark debris,
produced after DIS off an internal nucleon, with given polarization.
The application to the
spectator SiDIS reaction,
with the detection of a deuteron in the final state, represents a first
play-ground
of our approach.
As already mentioned,
besides its pedagogical relevance, spectator SiDIS reactions
could provide unique information on
the polarized EMC effect
and on the
hadronization mechanism of the quark debris during its propagation
through the nuclear medium, depending upon the kinematical region one chooses.
{ In particular, once a
final deuteron is detected, one can extract information on the proton. 
To play the same game for the neutron, one should
consider a polarized $^3$H target, and still  detecting a final deuteron. This SiDIS reaction could seem out of the present
range of experimental possibilities, but we cannot refrain to mention that valuable achievements have been reached in the last decade in dealing
with such a difficult target, as demonstrated by the final approval (with
scientific rating A)  of an experiment dedicated to  DIS by a
$^3$H target, at JLab \cite{marathon}.}
Finally, it should be pointed out that
the generalization to other SiDIS processes is rather straightforward,  even if
very heavy from the numerical-evaluation side, and it
will be presented elsewhere \cite{tobe}.

The paper is organized as follows. In  Section II we present the
basic formalism for the cross section,
valid for any   SiDIS process and define
the main quantities relevant for the calculations
within the PWIA framework.
In Section III, the spectator SiDIS reaction
 $\vec{^3{\rm He}}(\vec e,e'~^2{\rm H})X $ is
investigated in detail, i) introducing   the distorted spin-dependent spectral
function, that represents the main ingredient of our method for implementing
FSI effects,
through  a generalized eikonal approximation, and ii) adopting
 an effective cross section obtained from a model of
hadronization of the quark debris. In Section IV, numerical results for the
components of the spin-dependent spectral function, both in PWIA and with FSI
effects taken into account, are presented.
  Moreover,   asymmetries  are also shown in the
  most favorable kinematics
for gathering information on both i) the
  structure function $g_1(x)$ of a bound  nucleon
and ii) the hadronization
  process.
In Section V, conclusions are drawn and perspectives  presented.
\section{The cross section}
The differential cross section for the  generic
SiDIS process depicted in Fig. 1 can be written ~\cite{barone,scopetta} as
follows
\be
      \frac{d\sigma}{d\varphi_e dx_{Bj} dy }=
\frac{\alpha_{em}^2\ m_Ny}{Q^4}~
 L^{\mu\nu}(h_l) W_{\mu\nu}^{s.i.}(S_A,Q^2,P_h).
   \label{crosa-1}
   \ee
  where, for  incoming and outcoming electrons,
 $Q^2 =-q^2= -(k-k')^2 = \vec q^{\,\,2} - \nu^2=4 {\cal E}
{\cal E}' \sin^2(\theta_e/ 2)$ is the 4-momentum transfer (with
$\vec q = \vec k - \vec k'$, $\nu= {\cal E} - {\cal E}' $ and
$\theta_e\equiv \theta_{\widehat{\vec k  \vec k'}}$);
$y=\nu/{\cal E}$,
$x_{Bj} = Q^2/2m_N\nu $  the Bjorken scaling variable,   $m_N$
 the nucleon mass,  $\alpha_{em}$  the electromagnetic
fine structure constant and $P_h$  the detected-hadron 4-momentum.

The leptonic tensor $L_{\mu\nu}$ is an exactly calculable quantity
in QED. In the ultra relativistic limit it gets the form
 \be
L_{\mu\nu}(h_l)=2 \left [k_\mu k_\nu' +k_\mu' k_\nu -(k\cdot k') g_{\mu\nu}
+ ih_l \varepsilon_{\mu\nu\alpha\beta}~ k^\alpha~q^\beta\right]~  ~,
\label{lmunu}
\ee
where $ h_e $ is the helicity of the incident electron and
the Levi-Civita tensor
$\varepsilon_{\mu\nu\alpha\beta}$ is defined as  $\varepsilon_{0123}=-1$.
The semi inclusive (s.i.) hadronic tensor of the
target with polarization   four-vector $S_A$ { and mass $M^2_A=P^2_A$ } is defined as
\begin{eqnarray}
W_{\mu\nu}^{s.i.}(S_A,Q^2,P_h)=&&\frac{1}{4\pi M_A} {\sum\limits_{X}}
 \la  S_A, P_A|J_\mu|P_{h},X\ra  \la P_{h},X|J_\nu| S_A, P_A \ra\nonumber\\&&
(2\pi)^4\delta^4\left(P_A+q-P_X-P_{h}\right)~d\tau_X ~\frac{d{\bf P}_h}{2E_h(2\pi)^3}.
\label{wmunuA}
\end{eqnarray}
{   where the covariant normalization
$\la p| p'\ra =2E(2\pi)^3\delta\left( {\bf p-p}'\right)$ has been assumed and
$d\tau_X$ indicates the suitable phase-space factor for the undetected
hadronic state $X$.}
It should be pointed out that
 in Eq. (\ref{wmunuA}) the
integration over the  phase-space volume of the detected hadron, $h$, does not
have
to  be performed.

\subsection{PWIA}
\begin{figure}[!ht]        
   \includegraphics[scale=0.35]{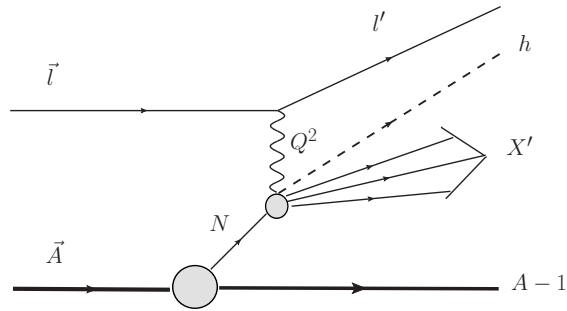}
\caption{
Diagrammatic representation of SiDIS processes in PWIA.
In   standard SiDIS reactions,
the hadron $h$, originated from the current quark fragmentation,
is detected. In spectator SiDIS processes, the
 $(A-1)$-nucleon system is detected { in place } of
the hadronic state  $h$.}
\label{Fig2}
\end{figure}

Within PWIA, the complicated final baryon states
$|P_{h},X\rangle$ are approximated by a
tensor product of hadronic states,  viz
$$|P_{h},X\rangle^{PWIA}=  |P_{A-1}\ra \otimes|P_h\ra \otimes |X'\ra$$
where  $|P_{A-1}\ra$ is { a short notation for indicating} the state
of the fully-interacting
$(A-1)$-nucleon  system,
which acts merely as a spectator, $|X'\ra$ the baryonic state, that originates
together with $|P_h\ra$ from the hadronization of the quark which
has absorbed the virtual photon, and of the other colored remnants.
Due to such an approximation of the final states, one can relate the nuclear
tensor $W_{\mu\nu}^{s.i.}(S_A,Q^2,P_h)$ to the one of  a single nucleon
$w_{\mu\nu}^{s.i.}(S_N,Q^2,P_h)$,
 { by performing the following steps: i) approximating
the nuclear current operator $J_\mu$ by
a sum of single nucleon operators
$j_\mu^N$ , ii)  disregarding the coupling of the
virtual photon with the spectator system (given the high momentum transfer),
 iii) neglecting for the present study the effects of the boosts
(that will
be properly taken into account in a Light-front framework elsewhere
\cite{tobe})
and  iv) inserting
in Eq. (\ref{wmunuA}) complete sets of nucleon plane waves and
$(A-1)$-nucleon   interacting states, given by
\be
\sum_\lambda \int {d {\bf P}_N\over 2 {\cal E}_N (2\pi)^3} ~ |\lambda,P_N \ra
 \la  \lambda, P_N|  ~= 1
 \nonu
\sum \! \!\! \!\! \!\! \!\int_{~\epsilon^*_{A-1}}
\rho\left(\epsilon^*_{A-1}\right)
~\int {d {\bf  P}_{A-1}\over 2 E_{A-1} (2\pi)^3} \, | \Phi_{\epsilon^*_{A-1}},{\bf P}_{A-1}\ra
 \la  \Phi_{\epsilon^*_{A-1}},{\bf P}_{A-1}|  ~= 1
\label{comp}
\ee
where $P_N\equiv \{{\cal E}_N=\sqrt{m^2_N+|{\bf P}_N|^2}, {\bf P}_N\}$,
$\Phi_{\epsilon^*_{A-1}}$ is the intrinsic part of the  $(A-1)$-nucleon
state, with  eigenvalue $\epsilon^*_{A-1}$, and
$E_{A-1}=\sqrt{(M^*_{A-1})^2+|{\bf P}_{A-1}|^2}$ with $M^*_{A-1}=Z_{ A-1} m_p+
(A-1-Z_{ A-1})m_n
+\epsilon^*_{A-1}$. The symbol with  the sum overlapping the
integral
 indicates that the $(A-1)$ system has both discrete and continuum energy spectra: this
 corresponds to negative and positive values of  the eigenvalue 
 $\epsilon^*_{A-1}$,
respectively. In Eq. (\ref{comp}), $\rho\left(
\epsilon^*_{A-1}\right)$ is the proper state density, that for  $A=3$ in the
two-body break-up (2bbu) and three-body break-up (3bbu)
reads
\be
\rho_{2bbu}= {1 \over (2 \pi)^3}~~,
\quad \quad \quad\quad \quad
\rho_{3bbu}= {1 \over (2 \pi)^6}~{m_N\sqrt{m_N\epsilon^*_2} \over 2}
\ee
In conclusion,
one obtains   the following expression of the nuclear tensor
\be
W_{\mu\nu}^{s.i.}(S^A,Q^2,P_h)=
{\sum\limits_{X',\lambda\lambda'}}
 \frac{1}{(2\pi)} \sum_N\int dE~ {\cal O}^{\hat {\bf S}_A}_{\lambda\lambda'}
({\bf p}_N,E)
 {\frac{1}{2E_N }}
\la \lambda',{ \tilde p}_N|j_\mu^N|P_h,X'\ra\la P_h,X'|j_\nu^N|\lambda,
 { \tilde p}_N\ra
\nonu \times
(2\pi)^4 \delta^4\left( P_A+q-P_{A-1}-P_h-P_{X'}\right)d\tau_{X'} \,
d{\bf P}_{A-1} \frac{d{\bf P}_{h}}{2E_h(2\pi)^3} ,
\label{munuN}
\ee
 where   i)  $P_{X'}+P_{A-1}$ is in place of $P_X$, ii) { the on-mass-shell
 four-momentum of the nucleon is}
 $ { \tilde p}_N\equiv \{E_N=\sqrt{m^2_N+|{\bf p}_N|^2},{\bf p}_N \}$ with
${\bf p}_{N}={\bf P}_A-{\bf P}_{A-1}$  the nucleon three-momentum, fixed
by the translational invariance of the
initial nuclear vertex
(c.f. Fig.\ref{Fig2}), viz
\be
\la  \Phi_{\epsilon^*_{A-1}},{\bf P}_{A-1}\lambda, \tilde  p_N|  S_A,  P_A\ra=
\sqrt{2E_N~2E_{A-1}~2M_A}~(2 \pi)^3~\times \nonu
\delta\left ( {\bf P}_{A}-{\bf P}_{A-1}-{\bf p}_{N}\right)~
\la\Phi_{\epsilon^*_{A-1}},{\bf P}_{A-1}\lambda, {\bf p}_N|  S_A,  \Phi_A\ra
\ee
where  $\Phi_A$ is the intrinsic wave function of the target nucleus, with mass
$M_A$ and the factor in front of the
delta function has been chosen in order to keep the notation of the intrinsic 
nuclear part as  close as possible
to the non relativistic case, where the plane waves have the normalization given by
$\la {\bf p}|{\bf p}'\ra= (2\pi)^3~\delta({\bf p}-{\bf p}')$. As a final remark,
let us remind   that $(P_A- P_{A-1})^2\ne m^2_N$. }
The effects of the nuclear structure in Eq. (\ref{munuN}) are encoded in
the overlaps
${\cal O}^{\hat {\bf S}_A}_{\lambda'\lambda}({\bf p}_N,E)$, defined as
 \begin{eqnarray}&& \!\!\!\!\!\!\!\!
{\cal O}^{\hat {\bf S}_A}_{\lambda\lambda'}({\bf  p}_N,E) = ~
\sum \! \!\! \!\! \!\! \!\int_{~\epsilon^*_{A-1}}\rho\left(
\epsilon^*_{A-1}\right)~
 \la  \Phi_{\epsilon^*_{A-1}},\lambda,{\bf p}_N|  S_A, \Phi_A\ra
 \la   S_A, \Phi_A| \Phi_{\epsilon^*_{A-1}},\lambda',{\bf p}_N \ra
 ~\times \nonumber \\
 && ~~~~~~~~~
 \delta\left( { E+ M_A-m_N-M^*_{A-1}-T_{A-1}}\right).
\label{overlaps}
\end{eqnarray}
{where $T_{A-1 }$ is the kinetic energy of the
 $A-1$ system. In a non
relativistic approach such contribution is disregarded,
leading to the identification of $E$ with
the usual missing energy, $E=\epsilon^*_{A-1}+B_A$, with $B_A$ the binding
energy of the target nucleus. It should be pointed out that
$m_N-E$ is the  energy
of a nucleon inside the target nucleus, where the $A-1$ system acts
as a spectator.}
 It is important to emphasize that
 the overlaps are nothing else but the matrix elements of the $2\otimes 2$ spin-dependent
spectral function of a nucleon inside the nucleus $A$,
with polarization ${\bf S}_A$ \cite{cda1},
 the crucial quantity to be introduced in the next section.
 The diagonal part yields
the probability distribution to find a nucleon in the
nucleus $A$ with three-momentum ${\bf p}_N$, missing energy $E$ and
 spin projection equal to $\lambda$. This entails  the following
 normalization
\be
{1 \over 2} \sum_\lambda \int dE~ \int  d{\bf  p}_N~
{\cal O}^{\hat {\bf S}_A}_{\lambda\lambda}( {\bf p}_N,E)=1~.
\ee

In what follows  we consider the polarized target in a pure state with
the nuclear wave functions having
definite spin projections on the spin quantization axis,   usually  chosen
along  the polarization vector ${\bf S}_A$. Accordingly,
in the complete set of the nucleon plane waves, the
spin projections $\lambda$ and  $\lambda'$ are defined
with respect to this direction.
As for the Cartesian coordinates, we  adopt the  DIS convention, i.e. the  $z$ axis
 is directed along the three-momentum transfer
$\bf q$ and   the plane $(x,z)$ is  the scattering plane. Notice that, in the DIS limit, the
direction of the three-momentum transfer coincides with that of the lepton
beam, ${\bf q}\ ||\ {\bf k}_e$.

{ Notice that   the semi-inclusive tensor defined
by Eq. (\ref{munuN}) refers to both kinds of SiDIS. Namely, the standard SiDIS
implies  integrations over $d\tau_{X'}$ and $ d{\bf P}_{A-1}$,
while for the spectator process the integrations are performed
over $d\tau_{X'}$ and $d{\bf P}_{h}/[2E_h(2\pi)^3]$, respectively.}
By inserting Eq. (\ref{munuN}) in Eq. (\ref{crosa-1}),
the cross section for standard SiDIS, when the hadron $h$ is detected,
is obtained as follows
\be
   2E_h~\frac{d\sigma(h_l)}{d\varphi_e dx_{Bj} dyd{\bf P}_h }
=\frac{\alpha_{em}^2\ y}{2Q^4}
  \ L^{\mu\nu}(h_l) \sum\limits_{\lambda\lambda'}\sum_N \int d{\bf p}_N \int
  dE~{m_N\over E_N}~
 w_{\mu\nu}^{s.i.}(p_N,P_h,\lambda\lambda')
{\cal O}^{\hat {\bf S}_A}_{\lambda\lambda'}({\bf p}_N,E)~,
  \nonu  \label{crosa-2}
   \ee
 where { the integration over ${\bf P}_{A-1}$ has been traded off with the one over ${\bf p}_N={\bf P}_A
 -{\bf P}_{A-1}$, and }
the  semi-inclusive  nucleon  tensor  (cf. Eq. (\ref{wmunuA})) is given by
\begin{eqnarray}
w_{\mu\nu}^{s.i.}(p_N,P_h,\lambda'\lambda)= \sum\limits_{X'}
\la  {  \tilde p}_N,\lambda'|j_\mu | P_h,X'\ra \la P_h,X'|j_\nu|
{  \tilde
p}_N,\lambda\ra
 \delta^4\left (  p_N+q-P_h-P_{X'}\right)d\tau_{X'}\, .
 \label{wm}
\end{eqnarray}
 where $ p_N =P_A -P_{A-1}\equiv \{m_N-E, {\bf
p}_N\}$ is such that $p^2_N\ne m^2_N={\tilde p}^2_N$.
The cross section for the spectator SiDIS
(when  the slow $(A-1)$ system is detected) has the same structure. However,
in this case, the integration over the hadronic variables $P_h$
has to be performed and the nucleon  tensor is of a pure inclusive DIS nature, viz   
\begin{eqnarray}
w_{\mu\nu}^{DIS}(p_N, Q^2,\lambda'\lambda)=\frac{1}{(2\pi)} \sum\limits_{
{  X''}}\la { \tilde p}_N,\lambda'|j_\mu |{  X''}\ra \la X''|j_\nu|{  \tilde p}_N,\lambda\ra
(2\pi)^4\delta^4\left ( p_N+q-P_{X''}\right)~d\tau_{X''}~.
\label{dis}
\end{eqnarray}
In Eq. (\ref{dis}), the final state $X''$ could be  $X'+h$, with the
notation in Fig. (\ref{Fig2}), but, obviously, could be any other state accessible
from the given initial state.
In this case, the cross section becomes   
\be
\frac{d\sigma(h_l)}{d\varphi_e dx_{Bj} dyd{\bf P}_{A-1} }
=\frac{\alpha_{em}^2\ y}{2Q^4}
  \ L^{\mu\nu}(h_l)  \sum\limits_{\lambda\lambda'}\sum_N
{  \int dE~\frac{m_N}{E_N} ~
w_{\mu\nu}^{DIS}(p_N,Q^2,\lambda'\lambda)
{\cal O}^{\hat {\bf S}_A}_{\lambda\lambda'}({\bf p}_N,E)}
\nonu \label{cross-3}\ee

  In conclusion, Eqs. (\ref{crosa-2}) and (\ref{cross-3}) show that
 the central
quantities for describing  SiDIS reactions in PWIA are i) the overlap integrals
${\cal O}^{\hat {\bf S}_A}_{\lambda\lambda'}({\bf p}_N,
E)$ which contain information
on the nuclear structure effects and ii) the
suitable  tensor $w_{\mu\nu}$ of a moving nucleon. In particular,
the antisymmetric part of the nucleon tensor is the basic
ingredient in the evaluation of proper cross section
asymmetries, that represent the main
 goal of the experimental investigation of SiDIS reactions.

\subsection{The    antisymmetric tensor $w_{\mu\nu}^{aDIS}$ of a moving nucleon }
  In this paper we focus on the spectator SiDIS, and in
particular on the asymmetries of the cross sections,  obtained properly varying the
polarization of the involved particles.
Therefore, the antisymmetric part of the nucleon tensor $w_{\mu\nu}^{DIS}$
(cf
 Eq.
(\ref{dis})) is the relevant quantity.
Following Ref. \cite{cda1}
 (see also Ref. \cite{SS}),
the antisymmetric part of the
tensor for a nucleon with a definite polarization $S_N$ is given by
\be
w_{\mu\nu}^{a,DIS}(p_N, Q^2,\lambda'\lambda)=\la \lambda'|\hat
w_{\mu\nu}^{aN}|\lambda \ra
\label{oper}\ee
where the  operator $\hat w_{\mu\nu}^{a N}(p_N,Q^2,{ S_N})$ in  Eq. (\ref{oper}) can be
written as
\cite{SS,cda1,antico,umnikovKapt}
\be
\hat w_{\mu\nu}^{aN}(p_N,Q^2,{ S_N}) =
i\varepsilon_{\mu\nu\alpha\beta} q^\alpha
\left [m_N
\hat S^\beta_N G_1^N({Q^2, p_N \cdot q})
\right.
\nonu +
\left.
\frac{G_2^N({ Q^2, p_N \cdot q})}{m_N}
\left(
(p_N \cdot q) \hat S^\beta_N - (S_N \cdot q) p^{\beta}_N\right)
\right ],
\label{op1}
\ee
where  the two scalar functions $G_{1,2}$  are
the polarized DIS structure functions. and the quantity $\hat S_N$
is the four-vector polarization operator acting in the
$2\times 2$ spin space. It is   defined as
\begin{eqnarray}
\hat S^\beta_N=\left \{
\begin{array}{lcc} \dfrac{\left (\boldsigma{\bf p}_N\right)}{m_N},
&& \beta=0 \\[2mm]
\boldsigma + {\bf p}_N\dfrac{\left (\boldsigma{\bf p}_N\right)}{m_N(E_N+m_N)},
&& \beta=1,2,3
         \end{array}\right. ,
         \label{operS1}
\end{eqnarray}
with $\boldsigma$  the usual Pauli matrices.

In standard SiDIS, the analogous of the operator
$\hat w_{\mu\nu}^{aN}(p_N,Q^2,{ S_N})$  becomes a more  complicated object, since,
within the quark parton model, it can be expressed
as a convolution of the TMDs
with  different quark fragmentation functions (see, e.g.,
Ref. \cite{barone}).

 \section{Spectator SiDIS by  a polarized $^3$He target}
 \label{ssidis}
  As shown in Eqs. (\ref{crosa-2}) and (\ref{cross-3}),
the nuclear effects in both  SiDIS reactions
 are governed by the overlap integrals
$ {\cal O}^{\hat {\bf S}_A}_{\lambda\lambda'}({\bf p}_N,E)$. In this paper we
focus mostly on the investigation of
nuclear effects, and therefore we consider the spectator SiDIS, that has
a nucleon tensor, { $w^{DIS}_{\mu\nu}$}, with less uncertainties in the parton structure
(calculations of the asymmetries for the standard SiDIS process
using different TMDs and fragmentation functions,
in the planned JLAB kinematics, will be reported elsewhere \cite{tobe}).
In particular, we consider the case of a polarized $^3$He target, { but
as mentioned
in the introduction, one can repeat the same considerations for $^3$H, modulo the
Coulomb effects.} For the sake of simplicity,  we choose the simplest channel,
 namely the one with a deuteron in the final
state,  this means that one can address the proton structure functions inside the
$^3$He target, while in the mirror nucleus one can study the neutron structure
functions. Generalization
to the case when   the detected system is a two-particle state
in the continuum  is straightforward, but more involved,
{ In particular,} we analyze
 polarized SiDIS with a longitudinal set up, i.e.
the polarization of the initial  electron and the target nucleus
are defined with respect to the direction of
the momentum transfer ${\bf q}$. { If the detected
unpolarized $(A-1)$-nucleon system is a deuteron,
 $\epsilon^*_{A-1}=-B_D$} and therefore
 the nucleon missing energy
is just the two body break-up (2bbu)
threshold energy of $^3$He, i.e. $E_{2bbu}=B_{^3\rm{He}}-B_D$.
For { the  final state we have chosen}, the cross section reads
\begin{eqnarray}
\frac{d\sigma^{\hat {\bf S}_A}(h_e)}
{{d\varphi_e}dx_{Bj} dyd{\bf P}_D }=
\frac{\alpha_{em}^2 \ { m_N} \ y }{ Q^4}
\ L^{\mu\nu}(h_e)
W_{\mu\nu}^{s.i.}(S^A,Q^2,P_D) ~
\label{crosa-3}
\end{eqnarray}
{ In the  asymmetry we are going to investigate,
 given by}
\be
\frac{\Delta \sigma^{\hat {\bf S}_A}}{d\varphi_e dx_{Bj} dy d{\bf P}_D}
\equiv\frac{ d\sigma^{\hat {\bf S}_A} (h_e=1) -d\sigma^{\hat {\bf S}_A}( h_e=-1) }
{d\varphi_e dx_{Bj} dy d{\bf P}_D},
\ee
only the antisymmetric part of both leptonic and nuclear tensors are
involved.
In particular, the antisymmetric part of the hadronic tensor
$W_{\mu\nu}^{s.i.}$ in Eq. (\ref{crosa-3}) reads
\be
W_{\mu\nu}^{a,s.i.}(S^A,Q^2,P_D)=\sum\limits_{\lambda\lambda'}
{ \frac{1}{2E_N}}
\la \lambda'|\hat w_{\mu\nu}^{aN}(p_N,Q^2,{ S_N})|\lambda\ra
{\cal O}^{\hat {\bf S}_A}_{\lambda\lambda'}(p_N,E_{2bbu})~.
\label{asymu}
\ee
with the nucleon DIS tensor $w_{\mu\nu}^{DIS}(p_N,Q^2,\lambda,\lambda')$ given
formally in Eqs. (\ref{dis}) and then
explicitly in Eqs. (\ref{oper}) and (\ref{op1}).
In the DIS limit,
the nucleon structure function $G_2^N$ yields, at the leading twist,
 a vanishing contribution to the measured cross section.
 Therefore,  in all the following  calculations,
contributions from $G_2^N$ are
neglected. Then, the antisymmetric part of the nucleon tensor becomes
\be
\la\lambda'|\hat w_{\mu\nu}^{aN}|\lambda\ra= iG_1^N(Q^2,p_N \cdot q)
\varepsilon_{\mu\nu\alpha\beta} { m_N} q^\alpha
\sum_\kappa(-1)^\kappa~\la  \lambda'|\sigma_{-\kappa}| \lambda\ra
 Tr \left(\frac12 \sigma_\kappa
\hat S_N^\beta\right ).
\nonu
=- i\sqrt{3}G_1^N(Q^2,p_N \cdot q)
\varepsilon_{\mu\nu\alpha\beta} { m_N} q^\alpha
\sum_\kappa(-1)^\kappa\la 1 -\kappa \frac12 \lambda|\frac12 \lambda'\ra ~
{\cal B}_\kappa^\beta
\label{ttt}
\ee
with \begin{eqnarray}
{\cal B}_\kappa^\beta\equiv Tr \left(\frac12 \sigma_\kappa \hat S_N^\beta\right ).
\label{drugoi}
\end{eqnarray}
Notice that  Eq. (\ref{drugoi}) defines a "double" vector  with double indices:
{  the  index
$\kappa=0,\pm 1$,    labels   three four-vectors},
 with Lorentz index $\beta$. {The latter
has to be contracted with the corresponding index in the    Levi-Civita
 tensor $\varepsilon_{\mu\nu\alpha\beta}$, see   Eq. (\ref{op1})}.
{  The Cartesian components of  ${\bm {\cal B}}^{\beta}$ are given by (a mixed
 notation is adopted, but self-explaining)
\begin{eqnarray}
  {\cal B}^\beta_i=\left \{
       \begin{array}{lcc}
       \dfrac{\left( {\bf p}_N\right)_i}{m_N}, && \beta=0 \\[2mm]
 \delta_{\beta i} +
 \left( {\bf p}_N\right)^\beta \dfrac{\left( {\bf p}_N\right)_i}{m_N(E_N+m_N)}, && \beta=1,2,3
         \end{array}\right  .
         \label{vecb}
\end{eqnarray}}
By placing Eq. (\ref{ttt}) into Eq. (\ref{asymu}), one can write
the nuclear tensor  as follows   
\begin{eqnarray}
W_{\mu\nu}^{a,s.i.}&\!\!(&\!\!   S^A,Q^2,P_D)= ~i{  {1 \over 2}} G_1^N(Q^2,p_N \cdot q)
\varepsilon_{\mu\nu\alpha\beta} {  { m_N\over E_N}} q^\alpha
\times\nonumber\\ && \times
\sum\limits_{\lambda\lambda'} \sum_{\kappa} (-1)^\kappa
\left[-\sqrt{3}\la 1 -\kappa \frac12 \lambda|\frac12 \lambda'\ra
{\cal O}^{\hat {\bf S}_A}_{\lambda\lambda'}({\bf p}_N,E_{2bbu}) \right]
~{\cal B}_\kappa^\beta
\label{prosto}
\end{eqnarray}
It can be seen that the dependence upon the index $\kappa$
 leads to
a scalar product of two vectors, viz
\begin{equation}
({\bm {\cal P}}^{\hat{\bf S}_A} \cdot {\bm  {\cal B}}^\beta)\equiv \sum_\kappa
(-1)^\kappa  {\cal P}_{-\kappa}^{\hat {\bf S}_A} {\cal B}^\beta_\kappa,
\end{equation}
where
\begin{eqnarray}
{\cal P}^{\hat{\bf S}_A}_\kappa({\bf p}_N,E_{2bbu})
\equiv -\sqrt{3}\sum_{\lambda\lambda'} \la 1 -\kappa \frac12 \lambda|\frac12 \lambda'\ra
{\cal O}^{\hat {\bf S}_A}_{\lambda\lambda'}({\bf p}_N,E_{2bbu})
\label{kappa}
\end{eqnarray} are the   spherical  components of
the vector ${\bm {\cal P}}^{\hat{\bf S}_A} $,
that represents the contribution to  the spin-dependent
spectral function from the polarization of the
target nucleus (see  Ref. \cite{cda1} for details)
in a pure state with polarization ${\bf S}_A$
(we reiterate that in Eq. (\ref{kappa})
the spin quantization is along the nuclear polarization ${\bf S}_A$).
Then,  the antisymmetric part
of the nuclear tensor reads   
\begin{eqnarray}
W_{\mu\nu}^{a s.i.}( S^A,Q^2,P_D)= ~i {  {1 \over 2}}G_1^N(Q^2,p_N \cdot q)
~\varepsilon_{\mu\nu\alpha\beta} { { m_N \over E_N}} q^\alpha
\left ({\bm {\cal P}}^{\hat{\bf S}_A}\cdot {\bm {\cal B}}^\beta\right)
\label{er}
\end{eqnarray}
{  For further   purposes, let us write more explicitly the components of
${\bm {\cal P}}^{\hat{\bf S}_A}$
 both in spherical and Cartesian coordinates.
 By using the  spherical versors, one has }
\begin{eqnarray}
{\bm {\cal P}}^{\hat{\bf S}_A}={\cal P}_{||}^{\hat {\bf S}_A} \  {\bf e}_0 +
 {\cal P}_{1\perp}^{\hat {\bf S}_A} \ {\bf e}_{+} +{\cal P}_{2\perp}^{\hat {\bf S}_A} \ {\bf e}_{-}
\label{razlozhenie}
\end{eqnarray}
where
${\bf e}_0 || {\bf  S}_A$ (see, also \cite{antico,cda1}) and
\begin{eqnarray}
\begin{array}{ccc}
\bf{e}_0=\left(  \begin{array}{c} 0\\ 0 \\ 1\\ \end{array}\right ),
&
\bf{e}_+=-\frac{1}{\sqrt{2}}\left( \begin{array}{c} 1\\i\\0\\ \end{array}\right ), &
\bf{e}_-=\frac{1}{\sqrt{2}}\left( \begin{array}{c}
                       1\\ -i\\0\end{array}\right),\\
\end{array}
\end{eqnarray}
or, in terms of Cartesian versors
\begin{eqnarray}
{\bm  {\cal P}^{\hat{\bf S}_A}}= {\cal P}_x^{\hat {\bf S}_A} \ {\bf e}_x +
{\cal P}_{y}^{\hat {\bf S}_A} \ {\bf e}_y+{\cal P}_{z}^{\hat {\bf S}_A} \  {\bf e}_z.
\label{cartezian}
\end{eqnarray}
 Usually, the  DIS kinematics is defined in a coordinate system  with
 the $z$-axis  along the three-momentum transfer $\bf q$,
 whereas the quantization direction to determine
 the particle polarizations  is along the beam direction $\bf k_e$.
 In the Bjorken limit ${\bf q}\simeq {\bf k}_e$ and the two
 directions coincide. { This remark will become helpful once FSI are
introduced.} The $x$-axis  is
then  chosen to be either in the scattering or in the reaction plane;
however,
${\bf e}_y=[{\bf e}_z\times {\bf e}_x]$.

In terms of the overlap integrals, Eq. (\ref{overlaps}), the
components of  ${\bm  {\cal P}}^{\hat{\bf S}_A}$  are expressed in spherical
basis by
\begin{eqnarray}
{\cal P}_{||}^{\hat {\bf S}_A}={\cal O}^{\hat {\bf S}_A}_{\frac12 \frac12}-
{\cal O}^{\hat {\bf S}_A}_{-\frac12 -\frac12};\quad
 {\cal P}_{1\perp}^{\hat {\bf S}_A}=-\sqrt{2}{\cal O}^{\hat {\bf S}_A}_{\frac12 -\frac12};\quad
 {\cal P}_{2\perp}^{\hat {\bf S}_A}=\sqrt{2}{\cal O}^{\hat {\bf S}_A}_{-\frac12 \frac12}~,
 \label{cyclic}
\end{eqnarray}
 and in Cartesian basis by
\begin{eqnarray}
 {\cal P}_z^{\hat {\bf S}_A}={\cal P}_{||}^{\hat {\bf S}_A};\quad  {\cal P}_x^{\hat {\bf S}_A}=
 2 \Re\  {\cal O}^{\hat {\bf S}_A}_{\frac12 -\frac12};\quad
 {\cal P}_y^{\hat {\bf S}_A}=-2 \Im \ {\cal O}^{\hat {\bf S}_A}_{\frac12 -\frac12}~,
\label{partial}
\end{eqnarray}

It should be noted that, since in Eq. (\ref{razlozhenie}) the last two terms
are mutually complex conjugated, one has only
only
two independent components, e.g., ${\cal P}_{||}^{\hat {\bf S}_A}$ and
${\cal P}_{1\perp}^{\hat {\bf S}_A}$. This is a consequence of the fact that
in the considered reaction  one  has
at   disposal only two vectors, ${\bf S}_A$ and
${\bf p}_N$ from which a pseudovector ${\bm {\cal P}^{\hat{\bf S}_A}}$
can be constructed \cite{cda1,lussino}, viz
\begin{eqnarray}
{\bm {\cal P}}^{\hat{\bf S}_A}({\bf p}_N,E)= {\bf S}_A B_1(|{\bf p}_N|,E) +
\hat {\bf p}_N \left ( \hat {\bf p}_N \cdot {\bf  S}_A
\right ) B_2(|{\bf p}_N|,E)~.
\label{salmesf}
\end{eqnarray}
{ where $B_1(|{\bf p}_N|,E)$ and $B_2(|{\bf p}_N|,E)$ are scalar functions
to be constructed from the overlaps.
}
 By using Eqs. (\ref{cyclic}) and some algebra, it is easily seen
that
Eqs. (\ref{razlozhenie}) and (\ref{salmesf}) { become }
 equivalent.
It should be emphasized  that in presence of FSI
the spin-dependent spectral function additionally depends  upon the direction of the
momentum transfer ${\bf q}$, so that
the simple form given in Eq. (\ref{salmesf}) does not longer hold.

{  Let us analyze in more details the 2bbu contribution to the
spin-dependent spectral function of a $A=3$ nucleus, within the PWIA framework.
In the  actual  calculations,
 both the  $^3$He (target) wave function and the deuteron one
correspond to exact solutions of the Schr\"odinger equation with
the AV18 nucleon-nucleon potential~\cite{av18}. In particular, for $^3$He,
the wave function  of  Ref.~\cite{pisa},
{ but without Coulomb effects, has been adopted, namely it can be applied for
describing also $^3$H.}

The overlaps ${\cal O}^{\hat {\bf S}_A}_{\lambda\lambda'}$ in Eqs.
(\ref{munuN}) and (\ref{overlaps}) are explicitly written as
\be
{\cal O}^{\hat {\bf S}_A}_{\lambda\lambda'}({\bf p}_N, E_{2bbu})=\sum_{M_D}\left[
\sum\limits_{\{ \alpha, \tilde\alpha\}}
\la XM_X L_\rho M_\rho|\frac12 M_A\ra \la \tilde X\tilde M_X \tilde L_\rho \tilde M_\rho|\frac12 M_A\ra
\la 1M_D \frac12 \lambda| X M_X\ra  \right. \nonu \left . \la 1M_D \frac12 \lambda'| \tilde X \tilde M_X\ra
(4\pi)^2 i^{L_\rho}(-i)^{\tilde L_\rho}
{\rm Y}_{L_\rho M_\rho}(\hat {\bf p}_N){\rm Y^*}_{\tilde L_\rho \tilde M_\rho}(\hat {\bf
p}_N) O_{\alpha}(|{\bf p}_N|,E_{2bbu})
 O_{\tilde \alpha}(|{\bf p}_N|,E_{2bbu}) \phantom{\dfrac12\!\!\!}\right ]
\nonu \label{oo}
 \ee
 where the radial overlaps $ O_{\alpha}(|{\bf p}_N|)$ are given by
 \begin{eqnarray}
 O_{\alpha}(|{\bf p}_N|,E_{2bbu})=\int  d\rho ~\rho^2  \int dr_{23}~ r_{23}~
j_{L_\rho}({\rho}|{\bf p}_N|) R_{\alpha}( r_{23},\rho) \Psi_{L_D}(r_{23})
\end{eqnarray}
  with  $\bfgr{\rho}$ and ${\bf r}_{23}$  the
two Jacobi coordinates: ${\bf r}_{23}={\bf r}_2 -{\bf r}_3$  and   ${\bfgr \rho}={\bf
   r}_1 -({\bf r}_2+{\bf r}_3)/2$. In Eq. (\ref{oo}),  $\{\alpha\}$ denotes
  the quantum numbers of a "deuteron-like"
configuration, i.e. $L_\rho, X, j_{23}=1, L_{23}=L_D=0,2$ with the corresponding
 projections
 $M_\rho, M_X, M_D$ (see below). Eventually $R_{\alpha}( r_{23},\rho)$ and
   $\Psi_{L_D}( r_{23})$ describe  the  target and deuteron radial
   wave functions, respectively, viz
   \be
   \langle \sigma_1,\sigma_2,\sigma_3; T_{23}, \tau_{23},
\tau; {\bfgr \rho},{\bf r}_{23}|^3{\rm He};
{1 \over 2} M_A;{1 \over 2} T_z\rangle =\nonu =\langle T_{23} \tau_{23} {1 \over 2}
\tau
|{1 \over 2} T_z \rangle~
\sum_{L_\rho M_\rho}
\sum_{X M_X} \sum_{j_{23}m_{23}}\la XM_X L_\rho M_\rho|{ 1 \over 2}M_A \ra
~ \la
j_{23}m_{23}{1 \over 2}\sigma_1|XM_X \ra
\nonu \times
\sum_{S_{23}m_{S_{23}}}
\sum_{L_{23} M_{23} } ~\la {1 \over 2}\sigma_2 {1 \over 2}
\sigma_3| S_{23}m_{S_{23}} \ra  \la L_{23} M_{23} S_{23} m_{S_{23}} |j_{23} m_{23} \ra
\nonu \times
{ Y}_{ L_{23} M_{23} } (\hat {\bf r}_{23})~ { Y}_{L_\rho M_\rho } (\hat {\bfgr \rho})
~\phi^{j_{23}L_{23} S_{23}}_{L_\rho X} ( r_{23},\rho)
 \label{hwf}\ee
 with $ \phi^{j_{23}L_{23} S_{23}}_{L_\rho X} ( r_{23},\rho)\equiv R_{\alpha}( r_{23},\rho)$
 , and
   \be
   \la {\bf r}_{23},| D; 1 M_D\ra= { \Psi_{0_D}(r_{23}) \over r_{23}}~{\cal
   Y}^{M_D}_{011}(\hat {\bf r}_{23}) + { \Psi_{2_D}(r_{23}) \over r_{23}}~
   {\cal Y}^{M_D}_{211}(\hat {\bf r}_{23})
  \label{dwf} \ee
    }
{  In Eq.~(\ref{oo}),  the $\phi$-dependence of   the  overlaps, and in turn of
 ${\bm {\cal P}}^{\hat{\bf S}_A}$ (cf
 Eq.~(\ref{partial}))
 is entirely governed by the difference $M_\rho - \tilde M_\rho$, which does not depend upon
 the internal summation}, namely  $M_\rho - \tilde M_\rho=\lambda'-\lambda$.
This implies, according to Eq. (\ref{cyclic}),  that the parallel
component ${\cal P}_{||}({\bf p}_N,E)$ does not depend upon $\phi$, while the perpendicular ones,
${\cal P}_{1\perp}({\bf p}_N,E)$ and ${\cal P}_{2\perp}({\bf p}_N,E)$, have a functional dependence given by
  $\exp(\mp i\phi)$,
respectively (see Ref.  \cite{antico}).
\subsection{Final state interaction effects}
\begin{figure}[!ht]
\includegraphics[scale=0.35 ,angle=0]{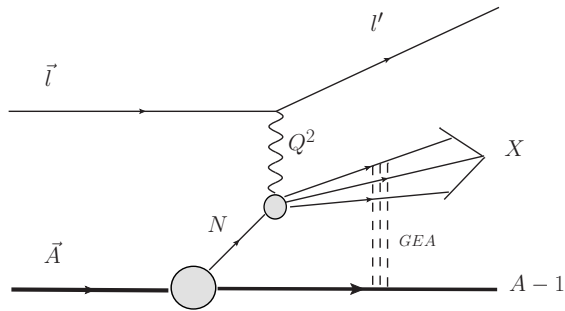}
\caption{{  A  diagrammatic illustration}  of FSI in spectator SiDIS.
The rescattering processes { between the quark debris and the nucleons inside the $(A-1)$ system} are treated within
a
generalized eikonal approximation (GEA)
\cite{ourlast}.}
\label{fsi}
\end{figure}

Let us now consider the effects of FSI.
They are due to i) the propagation of the
 nucleon  debris formed after  the $\gamma^*$ absorption by a target quark,
 followed  by its  hadronization  and ii) the interactions
 of the  produced  hadrons with the  $(A-1)$ {  spectator system}, as schematically depicted in Fig. \ref{fsi}.
Indeed, the calculation of such  FSI effects from first principle
represents  a very complicated many-body problem, so
that proper model approaches have to be developed.
To this end, one is guided by the observation that
in the  { DIS} kinematics we are considering: i) the momentum of the spectator
nucleus $|{\bf P}_{A-1}|$ is  small;
ii) the  large momentum transfer $|{\bf q}|$  leads to
  a very large relative momentum  between the
 debris (with momentum  ${\bf p}_X$)   and   the  nucleon  $i$ { (with momentum
  ${\bf k}_i$)} inside   the $(A-1)$  system,  i.e. $ |({\bf p}_X-{\bf k}_i)| \simeq |{\bf
  q}|\gg |{\bf k}_i |$ { (remind that the
  distribution of $|{\bf k}_i|$ is driven by the target wave function)};
   iii) the momentum transfer in the rescattering processes,  i.e. when the
debris interacts with { the nucleons inside} the $(A-1)$ system, has
    the typical magnitude of the  high-energy elastic $NN$ scattering, i.e. much smaller than the
   incident momentum  $ {\bf p}_X$ of the debris.
In this case, the rescattering  wave function
  can be   {   approximated by   its eikonal form   
  {  (in terms of T-matrix: $T({\bf p}_X,{\bf k}_i;{\bf p}'_X,{\bf k}'_i) \to T^{eik}({\bf
  p}_X$ )}, that allows one to describe the propagation of the
   debris produced  after the  $\gamma^*$ absorption by a target quark, while
   both   hadronization processes and interactions between
  the newly produced  pre-hadrons and the spectator nucleons take place}. This series of soft
 interactions with the spectator system can be characterized by an effective cross
 section  $\sigma_{eff}(z,Q^2,x)$ that  depends upon time
 (or the distance $z$ traveled  by the system $X$). {  Such an effective cross
 section allows one to construct a
 realistic profile function, that determines the eikonal approximation
 (see below)
\cite{ourlast,veronica,ckk,sigmaeff}}.
As a result,  {  in presence of FSI, the  PWIA overlaps given  in
Eq. (\ref{overlaps})
should be replaced by the suitable  ones that encode FSI effects}. In the 2bbu
channel, where   {  the asymptotic  three-momentum of the spectator system is
${\bf P}_{A-1}={\bf P}_D$}, one has,
{for the  matrix elements of the one-body current, an expression
{ that has the following schematic form }
\be
\la \hat S(1,2,3)\{ {\bf p}_X; {\bf P}_D \Psi_D \}|j_\mu^N|\Psi_{He}\ra =
\int d{\bf p}_{N}
\la {\bf p}_X|j_\mu^N |{\bf p}_{N}\ra ~\la{\bf p}_{N};{\bf P}_D \Psi_D|
\hat S_{Gl}(1,2,3) |\Psi_{He}\ra \ee
 where $\hat S_{Gl}(1,2,3)$, represents  the debris-nucleon eikonal
scattering S-matrix, that depends  upon the relative coordinates only, and it has been
assumed  to commute with $j^\mu $.
{ This  leads to consider  overlaps like $\la{\bf p}_{N};{\bf P}_D \Psi_D|
\hat S_{Gl}(1,2,3) |\Psi_{He}\ra $}
 (cf Eq. (6) in Ref. \cite{ourlast}). }
The operator $\hat S_{Gl}(1,2,3)$ can be written as follows
 \begin{equation}
\hat S_{Gl} ({\bf r}_1,{\bf r}_2,{\bf r}_3)=
\prod_{i=2,3}\bigl[1-\theta(z_i-z_1)\Gamma({\bf b}_1-{\bf b}_i,{ z}_1-{z}_i)
\bigr]~,
\label{SG}
\end{equation}
where  ${\bf b}_i$ and  $z_i$
are the perpendicular and parallel components of
${\bf r}_{i}$ (remind that ${\bf r}_1+{\bf r}_2+{\bf r}_3=0$), with
respect
to the direction of the propagation of the debris ${\bf p}_X$.
In the DIS limit
 ${\bf p}_X \simeq{\bf q} $ and the eikonal S-matrix
is defined with respect to $\bf q$.  {This implies that a dependence upon
${\bf q}$ has to be taken into account, but it is not explicitly indicated to avoid a too
heavy notation; however this will be recalled at the { proper} places.}
The profile function, $\Gamma$, is given by
\be
\Gamma({{\bf b}_{1i}},z_{1i})\, =\,\frac{(1-i\,\alpha)\,\,
\sigma_{eff}(z_{1i})} {4\,\pi\,b_0^2}\,\exp \left[-\frac{{\bf
b}_{1i}^{2}}{2\,b_0^2}\right]~.
 \label{eikonal}
\ee
where ${\bf r}_{1i}=\{{\bf b}_{1i}, {\bf z}_{1i}\}$
with ${\bf z}_{1i} ={\bf z}_{1}-{\bf
z}_{i}$  and ${\bf b}_{1i}={\bf b}_{1}-{\bf b}_{i}$.
It can be seen that,
 in the present generalized eikonal approximation,
unlike in the standard Glauber approach,
the profile function $\Gamma$ depends not only
upon the  transverse relative  separation
but also upon the longitudinal separation $z_{1,i}$ due to the $z$- (or time)
 dependence of the effective cross section $\sigma_{eff}(z_{1i})$ and the causal $\theta$-function,
  $\theta(z_i-z_1)$.
In principle, the effective cross section, $\sigma_{eff}(z_{1i})$
also depends on the total energy of the debris, $W^2\equiv P_X^2={ (p _N +q)^2}$.
{However, if the energy is not too large and the
hadronization process takes place inside the nucleus
$(A-1)$, the dependence on $W^2$ { is weak}, and
the number of produced hadrons can be taken constant.
}
Therefore, one can assume
$\sigma_{eff}(z_{1i},x_{Bj},Q^2)\sim \sigma_{eff}(z_{1i})$
\cite{sigmaeff,ourlast}. {  In conclusion, within the adopted approximation, the overlaps that include the FSI effects
are given by
\begin{eqnarray} &&
{\cal O}^{S_A(FSI)}_{\lambda\lambda'}({\bf P_D},E_{2bbu})=\nonumber \\ &&\left\la \hat S_{Gl}(1,2,3)\left\{
\Psi_{{\bf P}_{D}},\lambda,{\bf p}_N\right\}|   S_A, \Phi_A\right\ra
\left\la  \Phi_A,  S_A| \hat S_{Gl}(1,2,3)\left\{\Psi_{{\bf P}_D},\lambda',{\bf p}_N \right\}\right\ra=
\nonumber\\ &&
\sum_{M_D}
\left[
\sum\limits_{\{ \alpha, \tilde\alpha\}}
\la XM_X L_\rho M_\rho|\frac12 M_A\ra \la \tilde X\tilde M_X \tilde L_\rho \tilde M_\rho|\frac12 M_A\ra
\la j_{23}m_{23} \frac12 \lambda| X M_X\ra \la \tilde j_{23}\tilde m_{23} \frac12 \lambda'| \tilde X \tilde M_X\ra
\right .\nonumber\\ &&
\la l_{23}\mu_{23}1 M_S |j_{23}m_{23}\ra
\la \tilde l_{23}\tilde\mu_{23}\tilde 1 M_S|\tilde j_{23}\tilde m_{23}\ra
\la  L_{D} m_{L} 1  M_S |1 M_{D}\ra
\la \tilde L_{D}\tilde m_{L}\tilde 1  M_S|1 M_{D}\ra
\nonumber\\ &&
O_{\alpha}^{(FSI)}({\bf P}_{D},E_{2bbu})~
O_{\tilde\alpha}^{(FSI)}({\bf P}_{D},E_{2bbu})~,
\label{lp}
\end{eqnarray}
where
\be
O_{\alpha}^{(FSI)}({\bf P}_{D},E_{2bbu})= \int d\bs \rho \int d{\bf r}_{23}
{\rm e}^{i{\bf P}_D\bs \rho}
S_{Gl} ({\bf r}_{23},\bs\rho){\Psi_{L_D}(|{\bf r}_{23}|)\over |{\bf r}_{23}|}~
{\rm Y}_{L_D m_L}(\hat {\bf r}_{23})
\nonu \times ~{\rm Y}_{L_\rho M_\rho}(\hat {\bs \rho})
{\rm Y}_{l_{23} \mu_{23}}(\hat {\bf r}_{23}) ~
 R_{\alpha}(|{\bf r}_{23}|,|\bfgr{\rho}|) .
\label{overlapsFSI}
\ee
 with $S_{Gl} ({\bf r}_{23},\bs\rho)$  the non-singular part of the matrix elements of
 $\hat S_{Gl}(1,2,3)$ (remind that the adopted eikonal S-matrix is  diagonal in the Jacobi-coordinate  basis).
 }
A further issue is represented by the  fact that the direction  of the target polarization-axis, $\hat {\bf
k}_e$,
is not totally
 parallel to the the direction which determines the
eikonal $S$-matrix, i.e.
  $\hat {\bf p}_X$. Indeed,
in the Bjorken limit,
the momentum transfer ${\bf q}$ is almost parallel to the beam direction
${\bf k}_e$ so that in this case one can choose the quantization $z$-axis  along the beam direction and
perform calculations of FSI effects
within such a coordinate system, since $\hat {\bf
k}_e \simeq \hat {\bf q}\simeq \hat {\bf p}_X$. However, at finite values
of $|{\bf q}|$, the beam direction
differs from the direction which determines the
eikonal $S$-matrix. To reconcile the polarization axis and the eikonal approximation, one needs
to rotate the
target wave function from the quantization axis of the polarization ${\bf  S}_A$ to
the system with $z$-axis along ${\bf q}$,   namely
   
\be
\la \theta,\phi|\Psi_{^3He}\ra_{\hat{\bf  S}_A} = \la
\theta',\phi'|D(0,\alpha,0)|\Psi_{^3He}\ra_{\hat{\bf  q}}=\nonu
=\cos(\alpha/2)\
\la \theta',\phi'|\Psi^{{\cal M}=1/2}_{^3He}\ra_{\hat{\bf  q}}
+ \sin(\alpha/2)\ \la \theta',\phi' |\Psi^{{\cal M}=-1/2}_{^3He}\ra_{\hat{\bf  q}}
\label{he}
\ee
where the subscript indicate the direction of the $z$-axis
with $\cos \alpha= \hat{\bf  S}_A \cdot \hat{\bf  q}$.
In this case, the tensor $W_{\mu\nu}^{s.i.}(  S_A, Q^2,P_h)$ in Eq. (\ref{munuN})
is modified and reads as
\begin{eqnarray}
W_{\mu\nu}^{s.i.}(S_A, Q^2,P_h)=\cos^2(\alpha/2)\ W_{\mu\nu}^{\frac12\frac12}+\
\sin^2(\alpha/2)W_{\mu\nu}^{-\frac12-\frac12}+ \sin\alpha\ \left[\frac12\left(
W_{\mu\nu}^{\frac12-\frac12}+W_{\mu\nu}^{-\frac12\frac12}\right)\right]
\label{roty}
\end{eqnarray}
where $W_{\mu\nu}^{{\cal M M}'}$ are defined with respect to   the new axis,
 i.e. parallel to ${\bf q}$.
{  Then, introducing the following overlaps with quantization axis
$\hat{\bf q}$
\be{\cal O}^{{\cal M M'}(FSI)}_{\lambda\lambda'} \left({\bf P_D},E_{2bbu}\right )=
\nonu  \left\la \hat S_{Gl}(1,2,3)\left\{
\Psi_{{\bf P}_{D}},\lambda,{\bf p}_N\right\}|  \Psi^{\cal M}_A\right\ra_{\hat{\bf q}}
~\left\la  \Psi^{\cal M'}_A| \hat S_{Gl}(1,2,3)\left\{\Psi_{{\bf P}_D},
\lambda',{\bf p}_N \right\}\right\ra_{\hat{\bf q}}
\ee
and making use of their property} under complex conjugation, namely
\begin{equation}
 {\cal O}^{{\cal M M}'}_{\lambda\lambda'}({\bf P}_{A-1},E)=
 \left(-1\right)^{{\cal M}+{\cal M}'+\lambda+\lambda'}\left(
{\cal O}^{-{\cal M}-{\cal M}'}_{-\lambda-\lambda'}({\bf P}_{A-1},E)\right)^*~,
\label{prop}\end{equation}
it can be shown that
the contribution to the distorted spin-dependent spectral function
due to the  target polarization
takes the form
\begin{eqnarray}
{\bm {\cal  P}}^{\hat {\bf S}_A}_{(FSI)}=\cos \alpha {\bm {\cal  P}}^{\frac12\frac12}_{(FSI)}
+\sin\alpha {\bm {\cal   P}}^{\frac12-\frac12}_{(FSI)}~.
\label{newsf}
\end{eqnarray}
{where ${\bm  {\cal   P}}^{{\cal M} {\cal M}'}_{(FSI)}$ are evaluated with quantization axis $\hat {\bf q}$ and
the relations ${\bm  {\cal  P}}^{-\frac12-\frac12}_{(FSI)}=-{\bm {\cal  P}}^{\frac12\frac12}_{(FSI)}$ and
${\bm  {\cal   P}}^{-\frac12\frac12}_{(FSI)}= {\bm  {\cal   P}}^{\frac12-\frac12}_{(FSI)}~$
have been exploited (see Ref. \cite{cda1}).}
As it happens in PWIA,  ${\bm  {\cal  P}}^{{\cal M M}'}_{(FSI)}$ can be can be decomposed as follows
\begin{eqnarray}
{\bm {\cal P}}^{{\cal M M}'}_{(FSI)}={\cal P}_{||(FSI)}^{{\cal M M}'} \  {\bf e}_0 + {\cal P}_{1\perp (FSI)}^{{\cal M M}'}
\ {\bf e}_{+} +{\cal P}_{2\perp (FSI)}^{{\cal M M}'} \ {\bf e}_{-}~,
\label{razlozhenie1}
\end{eqnarray}
{where  ${\cal P}_{||(\perp)}^{\cal M M}$   are defined in  full analogy
with Eq. (\ref{partial}), while  ${\cal P}_{||(\perp)}^{\cal M -  M}$ are
given by }
\begin{eqnarray}&&
{\cal  P}_{||(FSI)}^{\frac12-\frac12}=\frac12\left[{\cal O}^{\frac12 -\frac12(FSI)}_{\frac12\frac12}
-{\cal O}^{\frac12 -\frac12(FSI)}_{-\frac12-\frac12} +c.c.\right]~,\nonumber\\&&
{\cal  P}_{1\perp(FSI)}^{\frac12-\frac12}=-\frac{1}{\sqrt{2}}
\left[{\cal O}^{\frac12 -\frac12(FSI)}_{\frac12-\frac12}
+{\cal O}^{*\ \frac12 -\frac12(FSI)}_{-\frac12\frac12}\right]~,\nonumber\\&&
 {\cal P}_{2\perp(FSI)}^{\frac12-\frac12}=\frac{1}{\sqrt{2}}
 \left[{\cal O}^{\frac12 -\frac12(FSI)}_{-\frac12\frac12}
+{\cal O}^{*\ \frac12 -\frac12(FSI)}_{\frac12-\frac12}\right]~,\label{kkk}
\label{nond}\end{eqnarray}
{In what follows, for the sake of brevity the diagonal components will be
indicated by only one projection, i.e.
${\cal P}^{\cal M M}({\cal O}^{\cal M M}) \to {\cal P}^{\cal M
}({\cal O}^{\cal M })$}.

\section{numerical results and discussion}

In this section,  numerical calculations of the  {  distorted spin-dependent } spectral
function in the 2bbu channel are presented. Particular attention
is paid to two typical  kinematics, known as parallel ($\hat {\bf p}_{N}
 \parallel
\hat z$, with
{\bf  $\hat z\equiv \hat {\bf q}$}) and perpendicular   ($\hat {\bf p}_{N}
\perp \hat z$) kinematics.
In the unpolarized case, the spectral function within these two kinematics
{  is influenced by} rather different physical {  effects}.
Namely, in the parallel kinematics, FSI are found to be negligibly small and,
accordingly, the process is suitable for  studying the DIS structure function of
a bound nucleon; differently, in the perpendicular kinematics the
 FSI effects  are predominant, so that details of the hadronization
 mechanism can be probed~\cite{ourlast}.  Bearing this in mind, let us consider
 the spectator SiDIS by a polarized target.  Usually, all quantities are
 presented in
 terms of {  the asymptotic three-momentum of the spectator system,
 ${\bf P}_{A-1}$}.
Moreover, to keep
the notation as close as possible to the one in the quasi elastic
$A(e,e'p)$-reactions, {  we introduce the missing momentum
 ${\bf p}_{mis}\equiv {\bf P}_{A-1}$ .}
 \begin{figure}\begin{center}
    \includegraphics[width=0.38\textwidth]{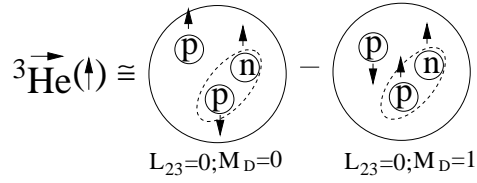}
  \caption{In a polarized $^3$He the total spin of the  proton pair is basically zero,
so that the neutron spin is mainly directed along the target polarization. The proton-neutron
spectator pair $"23"$ forms the deuteron with $L_{23}\sim 0$ in the final state. For an easy
representation, the contribution of the $D$-state is not depicted.}
\label{fig:pic3}
 \end{center}
\end{figure}
 Before going into the numerical analysis, let us have a qualitative
 glance at the intrinsic structure of a polarized $^3$He nucleus { (that represents
 our test ground)}. It is known that such a nucleus
 basically represents  a "polarized neutron".   As a matter of fact,
 in the polarized $^3$He the spin projections of the protons almost
 ($\sim 90\%$)
cancel each other, and the nuclear polarization is governed by that
of the neutron \cite{friar} (see  Fig.~\ref{fig:pic3}).
 This implies that, within PWIA {  when a deuteron acts as a spectator,}
the spin of the neutron in the final deuteron is expected
to be  directed along its initial polarization, i.e.
along the polarization of the  target.
Correspondingly,  the parallel {  component of the spin-dependent}
spectral function,
${\cal P}_{||}^{\frac12}={\cal O}^{\frac12}_{\frac12 \frac12}-
{\cal O}^{\frac12}_{-\frac12 -\frac12}$, gets
the main contribution  from the deuteron configurations with
$M_D=0$ and $M_D=1$. {  This can be easily understood considering only the deuteron
S-wave in  Eq. (\ref{lp}).  Indeed  putting
$L_\rho=L_D=l_{23}=0$, one can see that in
${\cal O}^{\frac12}_{\frac12 \frac12}$ the component with $M_D=0$  contributes
and  in
${\cal O}^{\frac12}_{-\frac12 -\frac12}$    the component  with $M_D=1$ acts (cf Eq. (\ref{dwf})).
Moreover, it turns out
 that  the contributions from  $M_D=0$ (with
 an upward neutron polarization) and  from $M_D=1 $
 have relative size
$1/2:1$, so that
${\cal P}_{||}^{\frac12} \simeq -\frac12\ {\cal O}^{\frac12}_{-\frac12 -\frac12}
$  and negative.}
Although the presence of  i) $P$- and $D$-waves in $^3$He and ii) the $D$-wave in
the deuteron
changes the simple scenario depicted in Fig.~\ref{fig:pic3}, at low missing momenta
one still expects that
${\cal P}_{||}^{\frac12} \simeq -\frac12\ {\cal O}^{\frac12}_{-\frac12 -\frac12}$.

\begin{figure}          
\parbox{8cm}{\includegraphics[width=8cm]{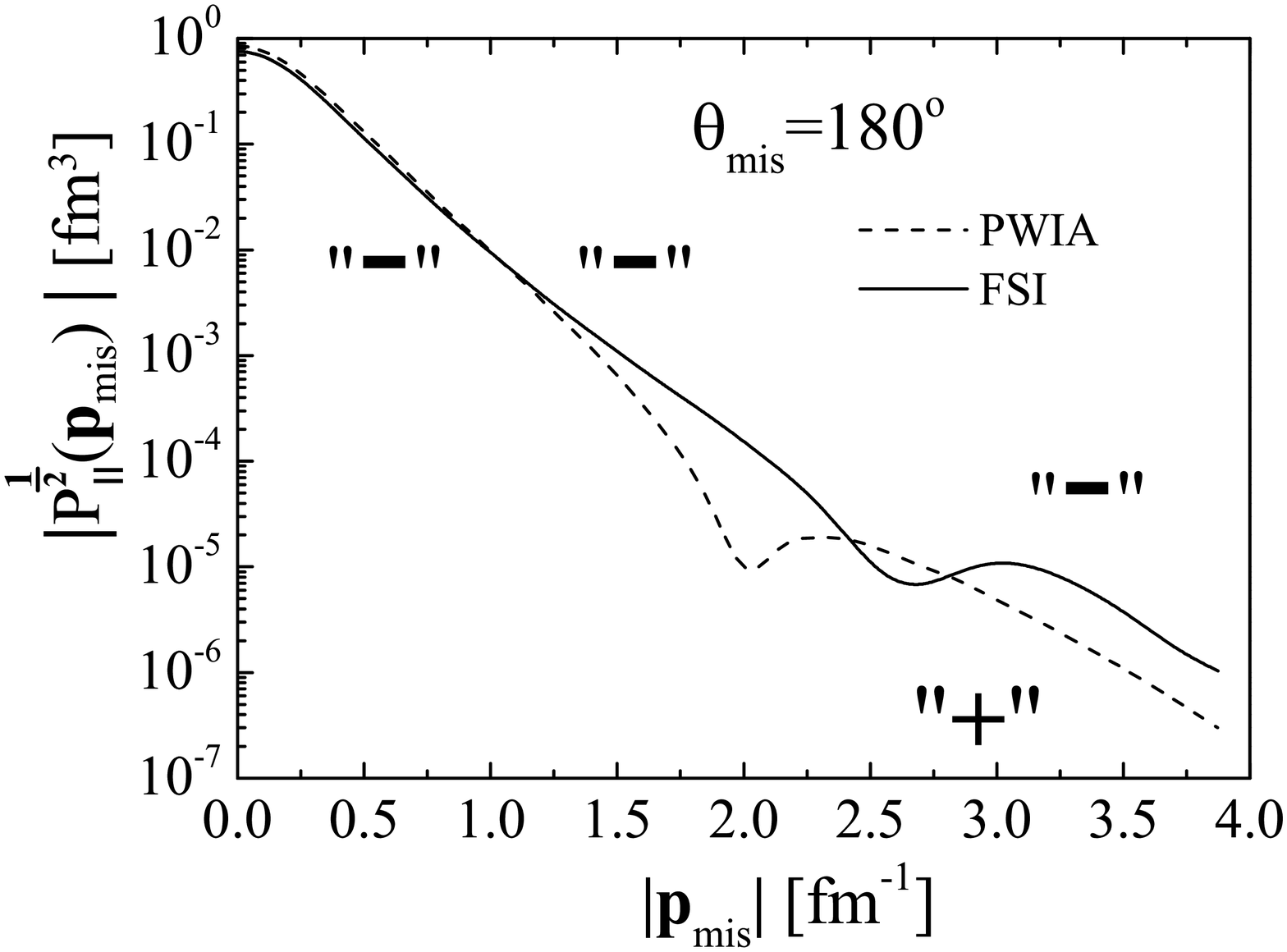}}
$~$\parbox{8cm}{\vspace{0.5cm} \includegraphics[width=8cm]{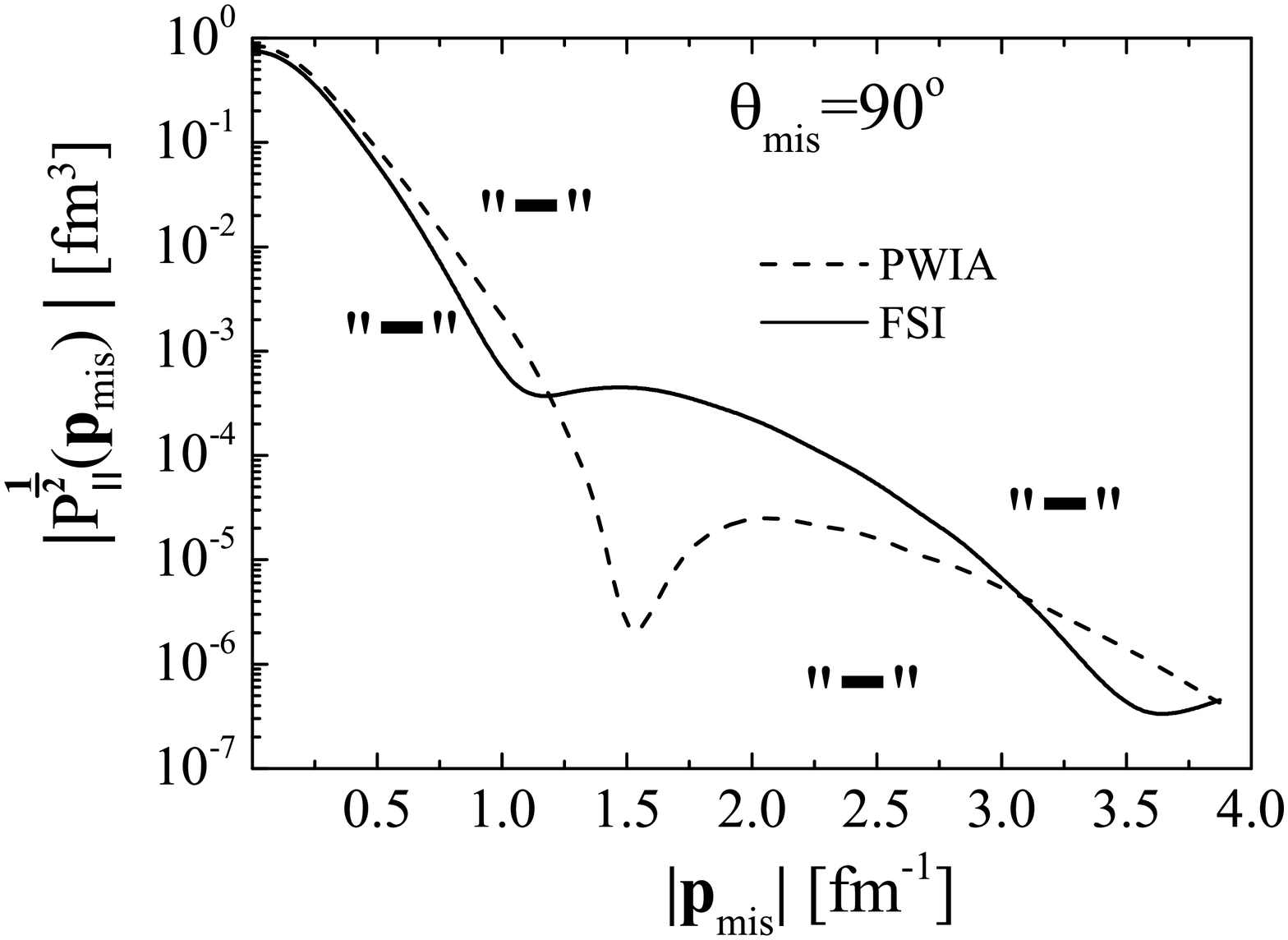}}
\caption{
{The absolute value { $\left|{\cal P}_{||}^{\frac12}\right|$}, { relevant
for a spectator SiDIS with a deuteron in the
final state}, for the
reaction $^3\vec{He} (\vec e,e'^{\ 2}\!H)X$, in the
Bjorken limit, vs the missing
momentum (${\bf  p}_{mis}\equiv{\bf P}_D$),
in parallel, $\theta_{mis}=180^o$ and $\phi_{mis}=180^o$  (left panel),
and perpendicular, $\theta_{mis}=90^o$ and $\phi_{mis}=180^o$ (right panel)
 kinematics.
Dashed line: PWIA calculations. Solid line: calculations with  FSI
 effects.
The inset symbols, {\large $+$} and {\large $-$}, indicate the sign of
{{${\cal P}_{||}^{\frac12}$}}.  Notice that
{{${\cal P}_{||}^{\frac12}$}}
with FSI effects  remains always negative, while
{{${\cal P}_{||}^{\frac12}$}} { in PWIA  changes sign only
in parallel kinematics}.}}
\label{figSparPerp}
\end{figure}

\begin{figure}
\includegraphics[width=0.48\textwidth]{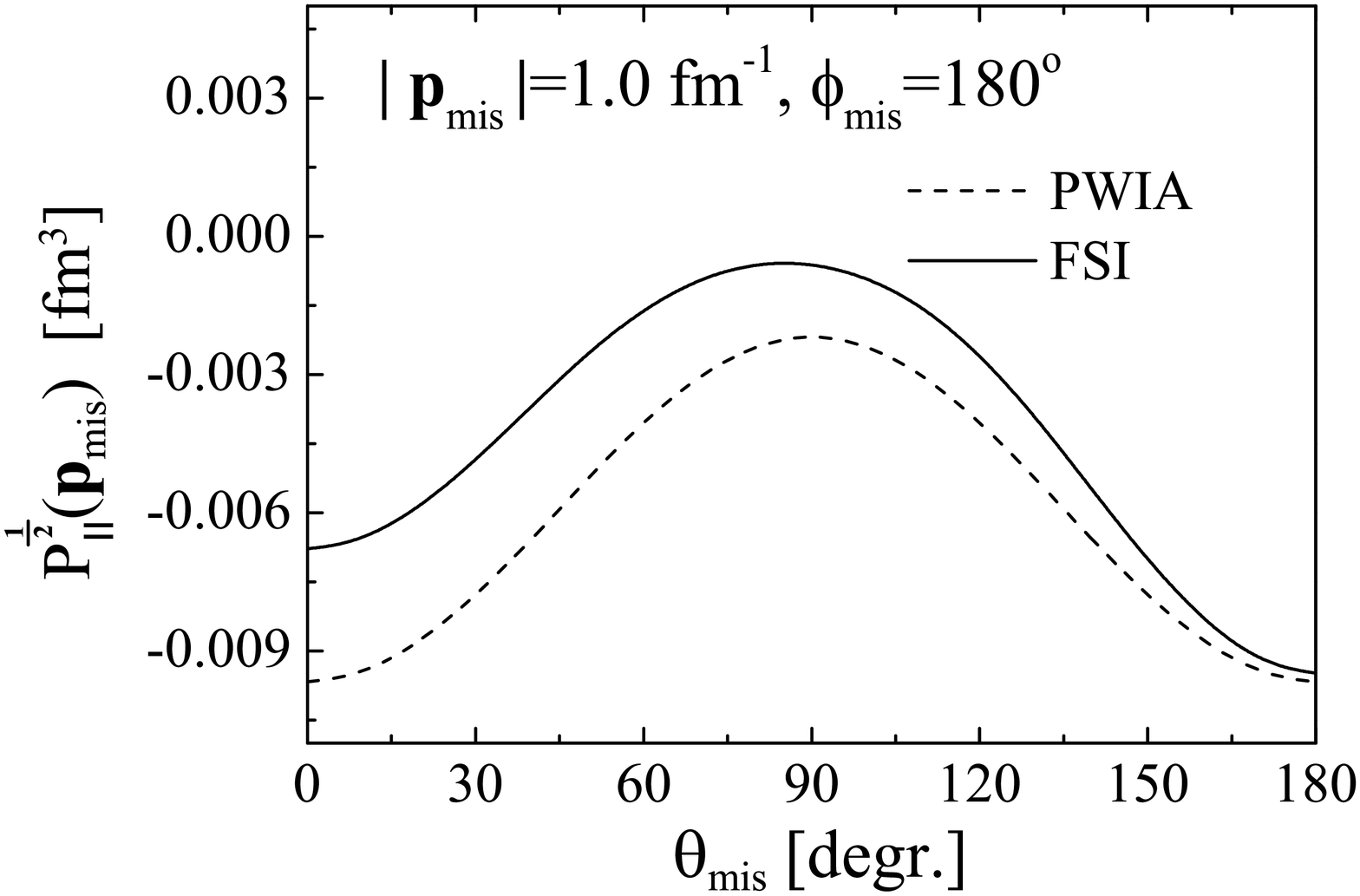}
\includegraphics[width=0.51\textwidth]{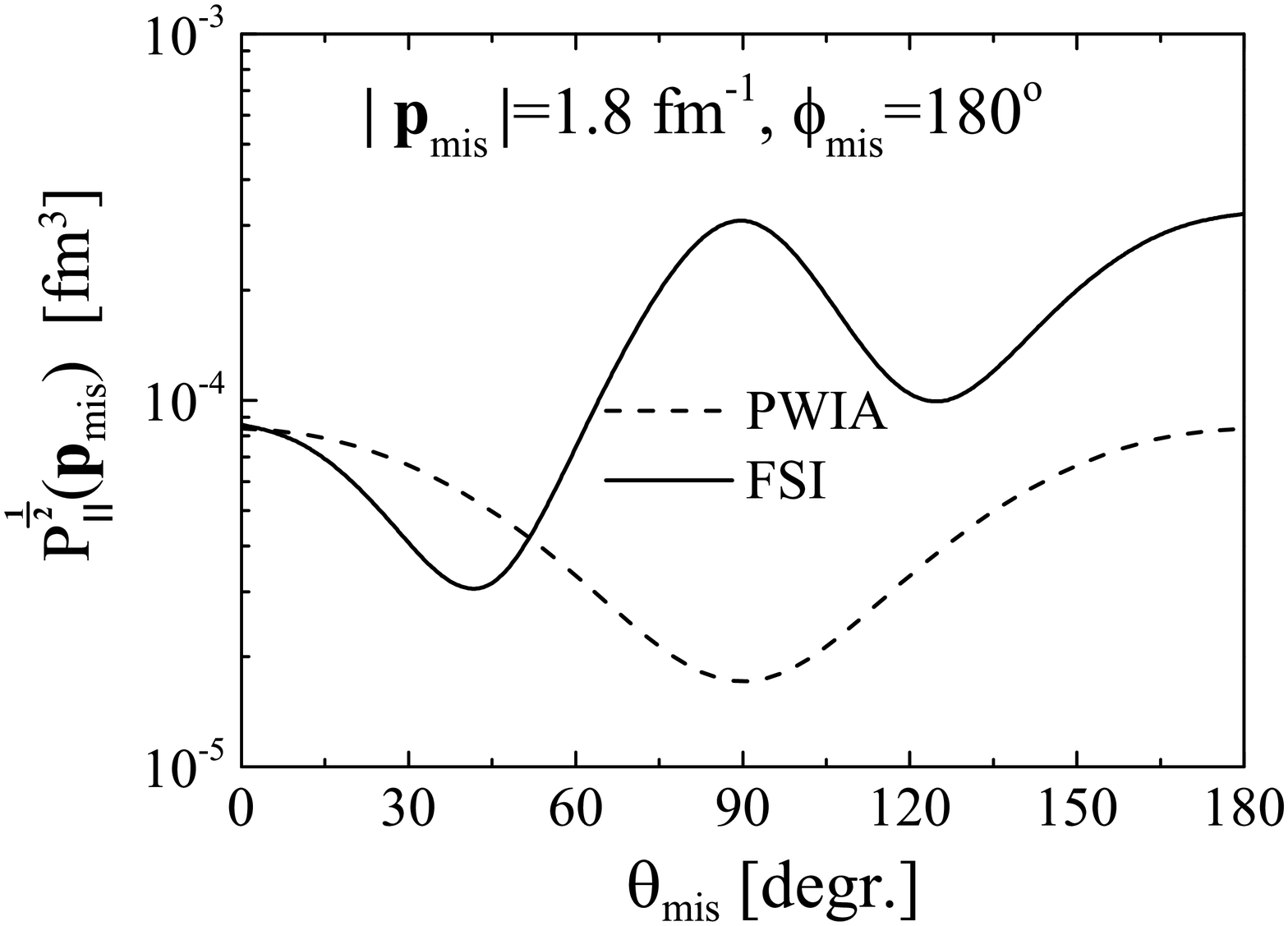}
\caption{The angular dependence   of
{ ${\cal P}_{||}^{\frac12}$},
for two  values of the missing momentum.
Dashed lines correspond to the PWIA calculations.
Solid lines include FSI effects.}
\label{spinDeOtugla}
\end{figure}

In Fig. \ref{figSparPerp}, {the absolute value of
  ${\cal P}_{||}^{\frac12 }$}
is shown  as a function of the missing momentum in both the parallel
($\theta_{mis}=180^o,\ \phi_{mis}=180^o$),
 and perpendicular ($\theta_{mis}=90^o$ and $\phi_{mis}=180^o$),
 kinematics. The dashed lines correspond to the
PWIA case, the
solid line is  { $\left|{\cal P}_{||}^{\frac12}\right|$} with  FSI effects included.
{  As mentioned above}, { ${\cal P}_{||}^{\frac12}$} at low missing momenta
is negative
 within both  kinematics, {  as indicated by the inset ''minus'' sign.}
   In the parallel kinematics {  (left panel)} at moderate values of
$|{\bf p}_{mis}|\sim 2 fm^{-1}$,   { ${\cal P}_{||}^{\frac12}$}
{  in PWIA } vanishes
and at higher
$|{\bf p}_{mis}| > 2 fm^{-1}$ becomes positive. This is an important feature
of the parallel component of the spin-dependent spectral function since,
as seen from Fig.~\ref{figSparPerp},
  { ${\cal P}_{||}^{\frac12 }$ with FSI effects }  never
   changes the sign, in both kinematics. This can be
{  exploited to determine}
the
presence (and strength) of FSI.
Notice  that, {  similarly to the unpolarized case~\cite{ourlast},
FSI are negligible at low values of $|{\bf p}_{mis}|$ (since in this case one
has a fast  final debris, given
${\bf p}_X \sim {\bf q}$) while
the FSI contribution becomes
sizable  for $|{\bf p}_{mis}| \geq 1 fm^{-1}$, where the  equal sign holds for the
perpendicular kinematics (right panel). Furthermore, because of the non trivial
angular dependence in ${\cal P}_{||}^{\frac12}$ (cf Fig.
\ref{spinDeOtugla}), though
in PWIA there be a zero at $|{\bf p}_{mis}|\sim \ 2 fm^{-1}$ in
parallel kinematics and  a minimum
at $|{\bf p}_{mis}|\sim \ 1.5 fm^{-1}$ in perpendicular kinematics,
the magnitude of FSI effects is much larger in this last setting.}
(cf. Fig. 4 of Ref.~\cite{ourlast}).

In Fig.~\ref{spinDeOtugla} we present the angular dependence
of  {{${\cal P}_{||}^{\frac12}({\bf p}_{mis})$}} for
fixed values of the missing momentum: $|{\bf p}_{mis}|=1 fm^{-1}$ (left panel) and
$|{\bf p}_{mis}|=1.8 fm^{-1}$ (right panel).  The choice $|{\bf p}_{mis}|=1 fm^{-1}$ has been inspired
by the fact that, as seen from Fig. \ref{figSparPerp}, FSI effects are still negligibly small
 (at least in the parallel kinematics), whereas $|{\bf p}_{mis}|=1.8 fm^{-1}$
corresponds to the region where the PWIA spectral function has a minimum,
 hence the FSI effects are
maximized.
It can be seen that, at lower missing momenta, FSI effects
are small  and   in
the backward hemisphere they can be safely neglected; at  $|{\bf p}_{mis}|=1.8 fm^{-1}$ the effects of
FSI are considerable, even predominant,  in the whole range of the missing angle.
{ It is worth noting that Figs.} \ref{figSparPerp} and \ref{spinDeOtugla}
can offer  hints  for choosing
the  { kinematics  for both spectator and standard SiDIS, in order to
minimize or maximize FSI effects. Let us remind that for the spectator SiDIS,
in the first case one can address the structure functions
of bound nucleons, and  in the second kinematics the hadronization mechanism
can be probed (see below).}

\begin{figure}
\includegraphics[width=0.47\textwidth]{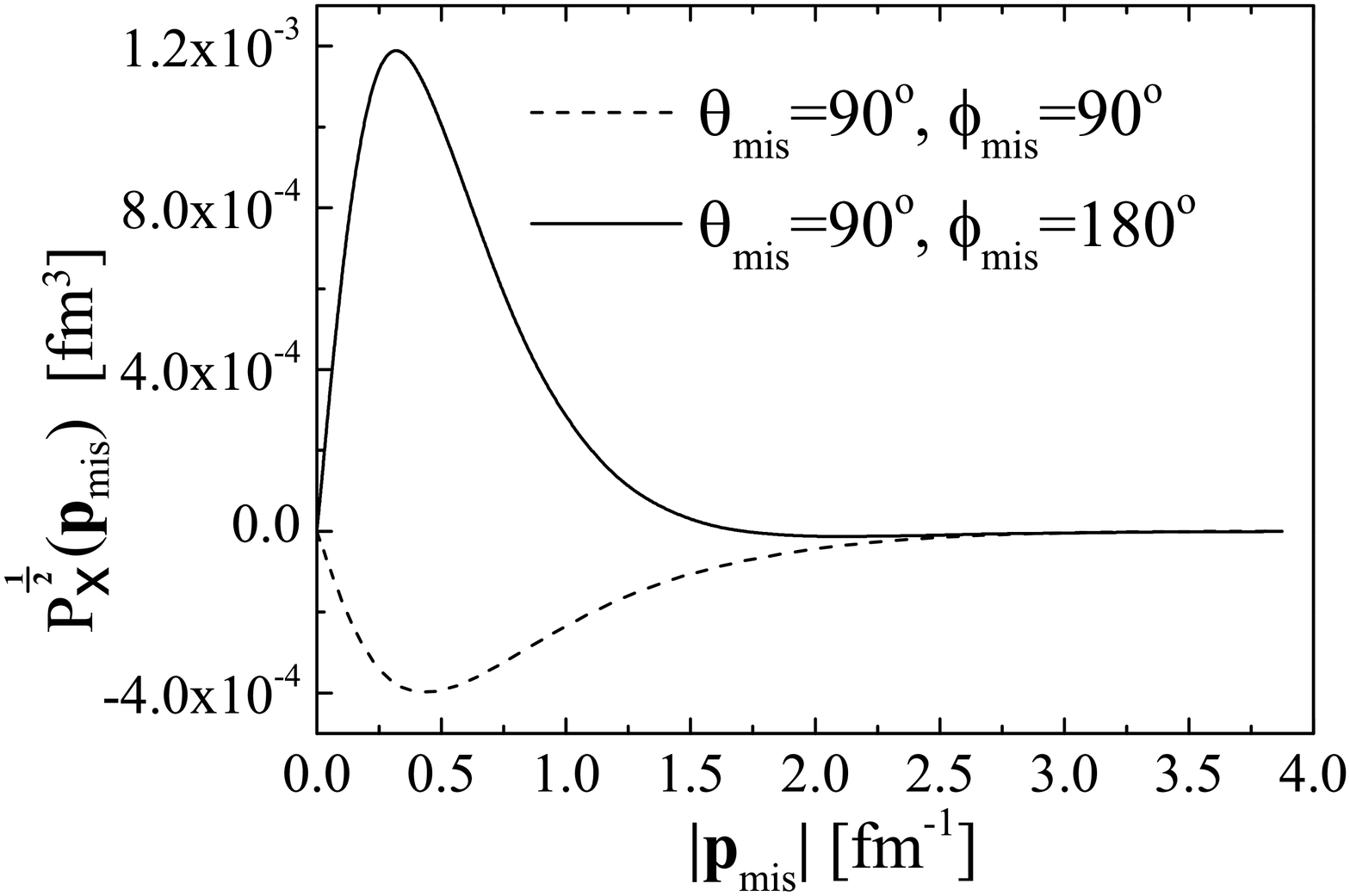}
\includegraphics[width=0.47\textwidth]{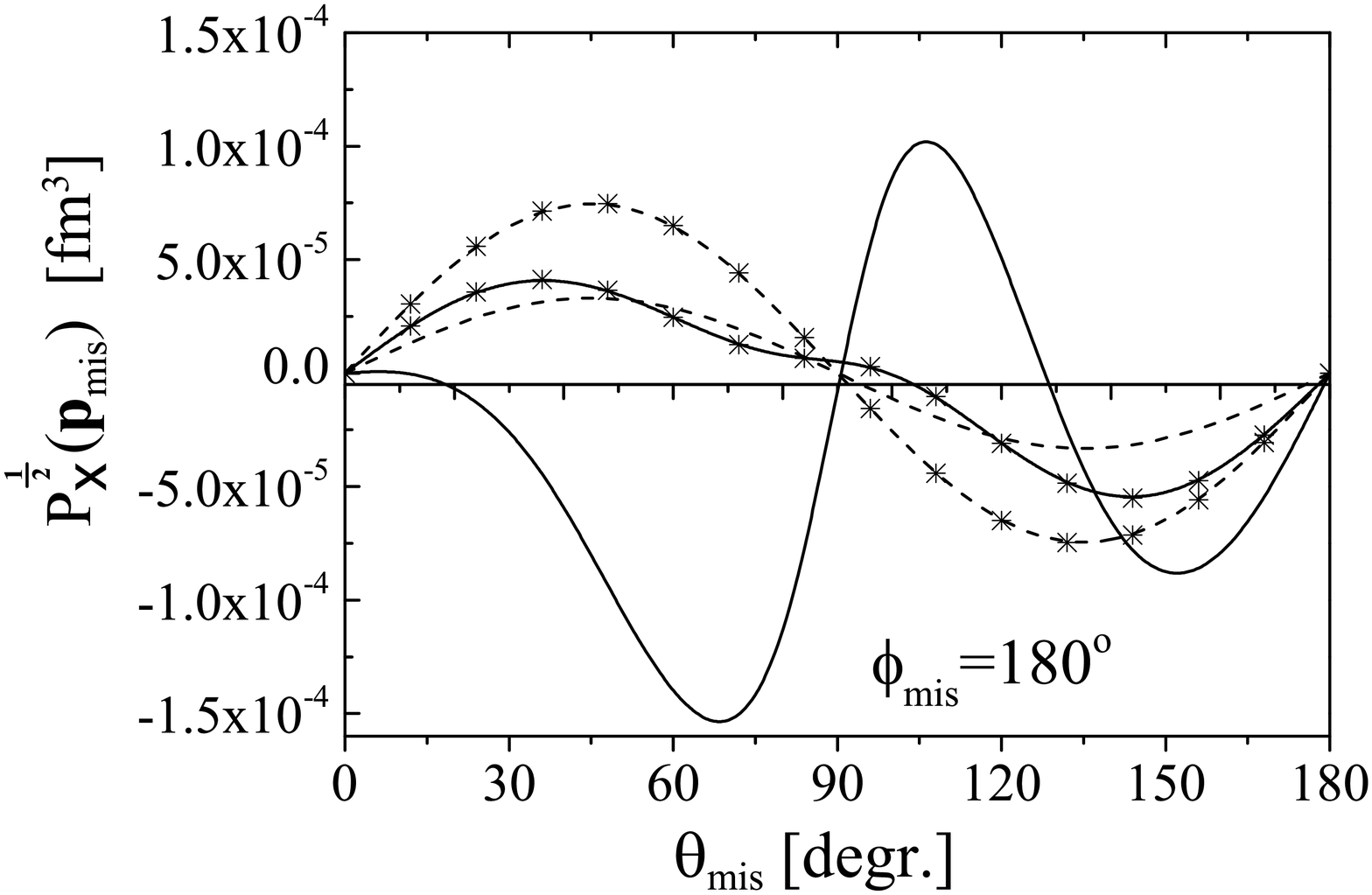}
\caption{
The $x$-component of the spin-dependent
spectral function, { relevant for a spectator SiDIS with a deuteron in the
final state}.
Left panel: ${\cal P}_x^{\frac12}({\bf p}_{mis})$ vs $|{\bf p}|_{mis}$,
in perpendicular kinematics,  $\theta_{mis}=90^o$, for
two values of  $\phi_{mis}$, $180^o$ (solid line) and
$ 90^o$ (dashed line).
Within such a kinematics the PWIA  spectral function is exactly zero, so that
the ${\cal P}_x^{\frac12}({\bf p}_{mis})$ is entirely due to the FSI effects.
Right panel:
angular dependence of   ${\cal P}_x^{\frac12}({\bf p}_{mis})$
for  two values of the missing momentum, $|{\bf p}_{mis}|=0.5 fm^{-1}$
(crossed lines)
and $|{\bf p}_{mis}|=1.8 fm^{-1}$. Dashed  lines:
  PWIA. Solid lines:
FSI effects are taken into account. Notice that, for a convenient presentation,
the results corresponding to
$|{\bf p}_{mis}|=0.5 fm^{-1}$ have been
rescaled by a factor $2\cdot 10^{-2}$.}
                \label{SpinPerp}
\end{figure}
Let us now briefly discuss the perpendicular components of the
spin-dependent spectral function, see  Eqs.~(\ref{cyclic}),(\ref{partial})
and (\ref{kkk}).  In the Bjorken limit, {  when $\hat {\bf q}$ becomes parallel
to the
$z$-axis, that in our analysis is also the  target polarization axis,
i.e.  $\hat z|| {\bf  S}_A$}, one can clearly see from  Eq.~(\ref{salmesf})
that $ {\cal P}^{\hat {\bf S}_A}_\perp$ in PWIA is exactly zero
within both the parallel (i.e. $\hat {\bf p}_N || {\bf
S}_A$) and perpendicular (i.e. $\hat {\bf p}_N \perp {\bf S}_A$)  kinematics,
(recall that only the term $\hat {\bf p}_N \left ( \hat {\bf p}_N \cdot {\bf  S}_A
\right ) B_2(|{\bf p}_N|,E)$ can contribute to the perpendicular
component of the spectral function, { since the term proportional
to ${\bf S}_A  B_1(|{\bf p}_N|,E)$ can
contribute only to the parallel one}).
\footnote{One should not confuse
the perpendicular and parallel kinematics, which refer to the direction
of nucleon momentum ${\bf p}_N$ with the parallel and perpendicular
components of the spectral function, which refer to the direction
of the vector ${\bm  {\cal P}}^{\hat {\bf S}_A}$}. But in presence of
 FSI, the spectral function
depends also upon the vector $\bf q$ (cf below Eq. (\ref{SG})), so that in Eq.~(\ref{salmesf})
  terms proportional to $\bf q$ must be included.
  In particular, a term like
  $\sim \hat {\bf p}_N \left( {\bf q}\cdot {\bf  S}_A \right)$ will contribute  in the perpendicular
  kinematics, causing ${\cal P}_\perp^{\hat {\bf S}_A}$ to  be different from zero.
  Therefore a nonzero value of  ${\cal P}_{1(2)\perp}^{\hat {\bf S}_A}$,
  in the perpendicular kinematics,
  undoubtedly points to  FSI effects.
Such a qualitative result can be  obtained in a more rigorous way by closely inspecting
 Eq.~(\ref{oo}), {  and investigating the dependence upon both $\phi_{mis}$
 and $\theta_{mis}$}.
It can be seen that the dependence upon $\phi_{mis}$ is determined by $M_\rho -\tilde M_\rho = \lambda '-\lambda$.
This  means that the parallel
spectral function does not depend at all upon $\phi_{mis}$, while the
$\phi_{mis}$
dependence of the perpendicular spectral function {will be
 ${\cal P}_{1(2)\perp}^{\hat {\bf S}_A}
\sim \exp(\pm i\phi_{mis})$}. Moreover, the presence of the term
$\sim {\rm Y}_{L_\rho M_\rho}(\hat {\bf p}_N){\rm Y^*}_{\tilde L_\rho \tilde M_\rho}(\hat {\bf p}_N)$
demonstrates  that ${\cal P}_{1(2)\perp}^{\hat {\bf S}_A}$ identically vanish
at
{$\theta_{mis}=0,~\pi/2, ~\pi$
 in PWIA, since $M_\rho -\tilde M_\rho =\pm 1$. As well-known,
 the two spherical harmonics can be expanded  on terms like  $\la L_\rho M_\rho  \tilde L_\rho
-\tilde  M_\rho|{\cal L} \pm 1\ra {\rm Y}_{{\cal L} \pm 1}(\hat {\bf p}_N)$
that vanish for
$\theta_{mis}=0, ~\pi$.
For $\theta_{mis}=\pi/2$  the  argument is less direct.
The two spherical harmonics are different
from zero only if $\tilde L_\rho +
\tilde M_\rho$ and $L_\rho +M_\rho$ are  both even,
 but $\tilde L_\rho +L_\rho$ is even and
 $\tilde M_\rho+M_\rho$ is odd.}
When the FSI effects are taken in to account, the previous product is
replaced with
$\int d\rho_\perp ....J_{\cal M}(|{\bf  p}_{mis\perp}|\rho_\perp)\cdot
\int d\tilde\rho_\perp ....J_{{\cal M}'}(|{\bf  p}_{mis\perp}|\tilde\rho_\perp)$,
where $J_{{\cal M}({\cal M}')}$ {  are the cylindrical Bessel functions
and one has still  ${\cal M}-{\cal M}'=\pm 1$. It is clear that,
in the parallel
kinematics (i.e. ${\bf p}_{mis\perp}=0$), { at most only one  Bessel
function cannot  vanish ($J_{\cal M}(0)\ne 0$ only for ${\cal M}=0$)}
 and therefore }
${\cal P}_\perp^{\hat {\bf S}_A}$ is zero
even in presence of FSI for ${\bf p}_{mis\perp}=0$.

One should notice  that {  for the spherical components  one has }
$ {\cal P}_{1\perp}^{\hat {\bf S}_A}= \left[{\cal P}_{2\perp}^{\hat {\bf
S}_A}\right]^*$, see  Eq. (\ref{cyclic}), but
 for numerical analysis it is more convenient to deal with
 real quantities, e.g. with the Cartesian components $ {\cal P}_x^{\hat {\bf S}_A}$  and  ${\cal
 P}_y^{\hat {\bf S}_A}$, see Eq.~(\ref{partial}). Since
 ${\cal P}_y^{\hat {\bf S}_A}(\phi_{mis})=-{\cal P}_x^{\hat {\bf S}_A}(\pi/2+\phi_{mis})$, it
 is sufficient to analyze only one component, say ${\cal P}^{\hat {\bf S}_A}_x$.
 In the left panel of Fig.~\ref{SpinPerp}, {  the Bjorken limit of }
 { ${\cal
 P}_x^{\frac12}(\phi_{mis})$ } is shown
 as a function of $|{\bf p}_{mis}|$, in the perpendicular kinematics and  two
 values of $\phi_{mis}$. It can be seen that { ${\cal P}^{\frac12}_x$},
 in comparison
 with ${\cal P}_{||}^{\frac12}$ (Fig. 6), is negligibly small,
  and in the perpendicular kinematics is entirely governed by FSI.
 In the right panel of Fig.~\ref{SpinPerp}, the angular dependence of
 ${\cal P}^{\frac12}_x$ , {  both without and with FSI effects},
  is presented for  two values of missing momentum. {  As already mentioned, the
  PWIA calculations vanish at $\theta_{mis}=0,~\pi/2,~\pi$.}
 Moreover, it is seen that the x-component is much smaller then the
 parallel ${\cal P}_{||}^{\frac12}$ in the whole range of $\theta_{mis}$.
 Finally, from Fig.~\ref{SpinPerp}, one could get the impression
that, at $\theta_{mis}=90^o$ the x-component of the spin-dependent
 spectral functions vanishes {  both without and with FSI effects}.
This is because of the adopted linear scale of the figure.
Actually, while in PWIA ${\cal P}_x^{\frac12}$ is exactly zero,
the calculations with FSI show that
 ${\cal P}^{\frac12}_x(\theta_{mis}  = 90^o) \sim 10^{-3}$.


\subsection{Cross sections and asymmetries}
In this subsection,  the relevance of different components
of the spin-dependent spectral
function is analyzed with respect to the application in
 {  spectator} SiDIS processes with polarized particles.
For the sake of brevity, in what follows  the notation
${\cal K}\equiv (k_e+k_e') $ is used.
In DIS limit, because of the scaling phenomenon, instead of the structure functions $G_{1,2}$  one introduces
 the more familiar scaling functions
$g_{1N}(x_N)=(p_N \cdot q)G_1^N(Q^2,p_N \cdot q)$
and
$g_{2N}(x_N) =(p_N \cdot q)^2/ m^2_N G_2(Q^2,p_N \cdot q)$,
{with $x_N=Q^2/2 q \cdot p_N$ } It can be shown that in the leading-twist
approximation the contribution of the
structure function $g_{2N}(x_N)$ to the cross section vanishes and, only
$g_{1N}$
is relevant to describe the antisymmetric part of the  tensor $w^{DIS}_{\mu\nu}$,
 Eq.~(\ref{op1}).
Therefore, the  contraction of the antisymmetric nuclear tensor
with the leptonic one results  in   
\begin{eqnarray}&&
L^{\mu\nu}(h_e)\ W^{a,s.i.}_{\mu\nu}( S_A,Q^2, P_D){  \sim}-2h_e
 {  {m_N\over 2 E_N}} Q^2 G_1^N(Q^2,p_N \cdot q){\cal K}^\beta \left ({\bm  {\cal B}}_\beta\cdot
 {\bm  {\cal P}}^{\hat {\bf S}_A}\right)\equiv
\nonumber\\ &&
={  h_e {Q^2 m_N \over E_N~q\cdot p_N}}  ~g_{1N}(x_N)
\left[\left( \KE
\cdot {\bm  {\cal P}}^{\hat {\bf S}_A}\right)+
\frac{\left( \KE  \cdot {\bf p}_N\right) }{m_N}
\frac{\left( {\bf p}_N \cdot {\bm {\cal P}}^{\hat {\bf S}_A}\right)}{E_N+m_N}-
{\cal K}_0\frac{\left( {\bf p}_N \cdot
{\bm {\cal  P}}^{\hat {\bf S}_A}\right)}{m_N}\right].
\end{eqnarray}

{ In order to experimentally single out    the spin-dependent part of the
cross section, one measures asymmetries of the cross sections
corresponding to the scattering of electrons with opposite helicities}.
In particular, the following asymmetry
   \begin{eqnarray}
&&\frac{\Delta \sigma^{\hat {\bf S}_A}}{d\varphi_e dx_{Bj} dy
{ d{\bf P}_{D}}}
\equiv\frac{ d\sigma^{\hat {\bf S}_A} (h_e=1) -
d\sigma^{\hat {\bf S}_A}( h_e=-1) }{d\varphi_e dx_{Bj} dy
{ d{\bf P}_{D}}} =
\nonumber\\ && =
  {   4} \frac{\alpha_{em}^2}{Q^2 z_N{\cal E}}
{m_N\over E_N}~g_{1N}\left (\frac{x_{Bj}}{z_N}\right )
   \left[\left( \KE \cdot {\bm  {\cal P}}^{\hat {\bf S}_A}\right)+
   \frac{\left( \KE \cdot {\bf p}_N\right) }{m_N}\frac{\left( {\bf p}_N \cdot
    {\bm  {\cal P}}^{\hat {\bf S}_A}\right)}{E_N+m_N}-
   {\cal K}_0\frac{\left( {\bf p}_N \cdot {\bm  {\cal P}}^{\hat {\bf S}_A}\right)}{m_N}\right],
   \label{crosa-21}
   \end{eqnarray}
allows one to single out  the spin-dependent part of the cross section in
 {  the
spectator SiDIS with a detected deuteron. In Eq. (\ref{crosa-21}), we
intentionally left  the subscript $N$,
since it can be applied to both  $A=3$ polarized  targets: $^3$He and $^3$H.
However, the actual calculations have been performed for  $\vec{^3{\rm He}}$
and therefore $N$ has to be substituted by  $p$, i.e. $ g_{1N}\to g_{1p}$.}
In Eq. (\ref{crosa-21}) $z_N$
is proportional to the light-cone
fraction of the momentum carried by the  nucleon in the nucleus,  i.e.
$z_N=(p_N\cdot q)/m_N \nu\approx 1-{p_{N3}}/{m_N}\approx A p^+_N/P^+_A$. {
It yields a measure of the difference between $x_{Bj}$ and $x_N$, that is the
actual variable upon which $g_{1N}$ depends.}
The asymmetry in  Eq.~(\ref{crosa-21})  has a factorized form, where
  the electromagnetic part of the interaction (the spin-dependent structure function $g_{1N}$)
is separated from the nuclear-structure effects (the term inside the  square
 brackets, proportional to the     spin-dependent spectral function
 ${\bm {\cal P}}^{\hat {\bf S}_A}$) and some kinematical factors.
 Such a form shows that in principle one can experimentally
 {  achieve a direct
 access to }  the
 spin-dependent structure function of a bound nucleon,
 provided that the expression within the square brackets can be  reliably calculated.
 Indeed, in  such a term, the nuclear part is {intertwined with}
 the electron kinematics, i.e.
the initial energy ${\cal E}$ and the energy transfer
 $y=({\cal E}-{\cal E}')/{\cal E}$,   but, fortunately, without
 a dependence upon
$x_{Bj}$. This favorable circumstance allows one  to vary the
kinematical variables { upon which $g_{1N}$ depends independently from the
ones of  the
nuclear-structure ${\bm  {\cal  P}}^{\hat {\bf S}_A}$}, as discussed  below.

 { For the sake of concreteness, let us consider how to extract
   information on the  bound proton by using
 a $^3$He target in pure spin state,} polarized
along the incident electron beam
(remind that in the Bjorken limit it coincides with $\bf q$
and generalization to arbitrary
polarization is straightforward, see Eqs.~(\ref{he})-(\ref{newsf})).
{ This  represents the   {\em longitudinally polarized}  target setting. The
transversely-polarized setting is quite important in the standard SiDIS, as
briefly discussed in the following subsection and in more detail in
 \cite{tobe}}.
For a reliable  study of the structure function { $g_{1p}$ } in the
spectator SiDIS, with a detected deuteron,
one has to  minimize uncertainties induced by both the nuclear structure terms and
FSI effects. To this end, we
fix the kinematics related to the spectral function, i.e.
${\bf p}_{mis}$, in such a way that FSI effects
are minimized, while we  vary
$x_N$ { (possibly
taking $z_N$ constant)}.
As clearly seen from Figs.~(\ref{figSparPerp}), (\ref{spinDeOtugla}) and
(\ref{SpinPerp}),
the most appropriate choice  for this purpose
is the parallel kinematics, $\theta_{mis}=180^o$ and
$|{\bf p}_{mis}|\simeq 1fm^{-1}$.  Within such conditions, one can safely
discard the contribution containing
 either
the perpendicular components of the spectral function
${\cal P}_{\perp}^{\hat {\bf S}_A}$
 or terms proportional to
 {   $|{\bf p}_N|^2/2m_N^2=|{\bf p}_{mis}|^2/2m_N^2<< 1$
 (within our approximation, recall that ${\bf p}_N= -{\bf p}_{mis}$).
  Then, for ${\bf p}_{mis}\parallel {\bf k}_e$
 one can approximate
the term inside the  brackets in Eq.~(\ref{crosa-21}) as
follows}
\be
\left[\phantom{\frac12}\cdots \right] \approx {\cal P}_{||}^{\frac12
 }({\bf p}_{mis}) {\cal E} {  \left [ 2-y-{Q^2 \over 2 {\cal E}^2}+
{ (2-y)}\frac{|{\bf p}_{mis}|}{m_N}
-{|{\bf p}_{mis}|^2 \over 2m_N^2} \right] }\approx
\nonu
\approx {\cal P}_{||}^{\frac12
}({\bf p}_{mis}) {\cal E} { (2-y)} \left [1
+\frac{|{\bf p}_{mis}|}{m_N}\right].
\label{kw}
\ee
where   ${\cal P}_{||}^{\frac12 }
({\bf p}_{mis}) \equiv \hat{\bf p}_N \cdot  {\bm {\cal P}}^{\frac12  }
({\bf p}_{mis})$
 and \be
 \KE \cdot {\bm {\cal P}}^{\frac12}({\bf p}_{mis})=
\left({\cal E} + {\cal E}' \cos \theta_e \right){\cal P}_{||}^{\frac12
 }({\bf p}_{mis})+ {\cal E}' \sin\theta_e {\cal P}_{\perp}^{\frac12
 }({\bf p}_{mis})=
\nonu \sim \left[{\cal E}-{Q^2 \over 2 {\cal E}} +{\cal E}'\right]
{\cal P}_{||}^{\frac12
 }({\bf p}_{mis})={\cal E}
\left[2-y-{Q^2 \over 2 {\cal E}^2} \right]{\cal P}_{||}^{\frac12
 }({\bf p}_{mis})
\nonu \nonu
\KE \cdot  {\bf p}_N=-  ({\cal E} +
{\cal E}' \cos \theta_e)~|{\bf p}_{mis}|= - {\cal E}
 \left(2-y -{Q^2 \over 2 {\cal E}^2}
\right)~|{\bf p}_{mis}|~~~.
\ee
In order to have a "pure" factorized form of the corresponding expressions, let us
define the reduced  asymmetry {  in parallel kinematics} as follows
\begin{eqnarray}
&&\Delta\sigma_{\|}\equiv
\frac{\Delta \sigma^{\frac12}(\theta_{mis}\sim 180^o)}
{d\varphi_e dx_{Bj} dy d{\bf P}_{D}}\left/
\left( { 4}{\frac{(2-y)~\alpha_{em}^2 \ { m_N}}{Q^2~E_N}}\right)=
g_{1N}\left (\frac{x_{Bj}}{z_N}\right )
 {\cal P}_{||}^{\frac12  }({\bf p}_{mis})\right . .
\label{tag}
\end{eqnarray}
It is seen that the reduced asymmetry factorizes into two terms: one entirely
determined by the structure function of the bound nucleon
$g_{1N}$, { that in the present calculation is $g_{1p}$},  the second
term being of a pure nuclear structure origin. We reiterate that
Eq.(\ref{tag}) has been obtained in the Bjorken limit
 and at  low values of the missing momenta  within the parallel kinematics.   

\begin{figure}
        \includegraphics[width=0.45\textwidth]{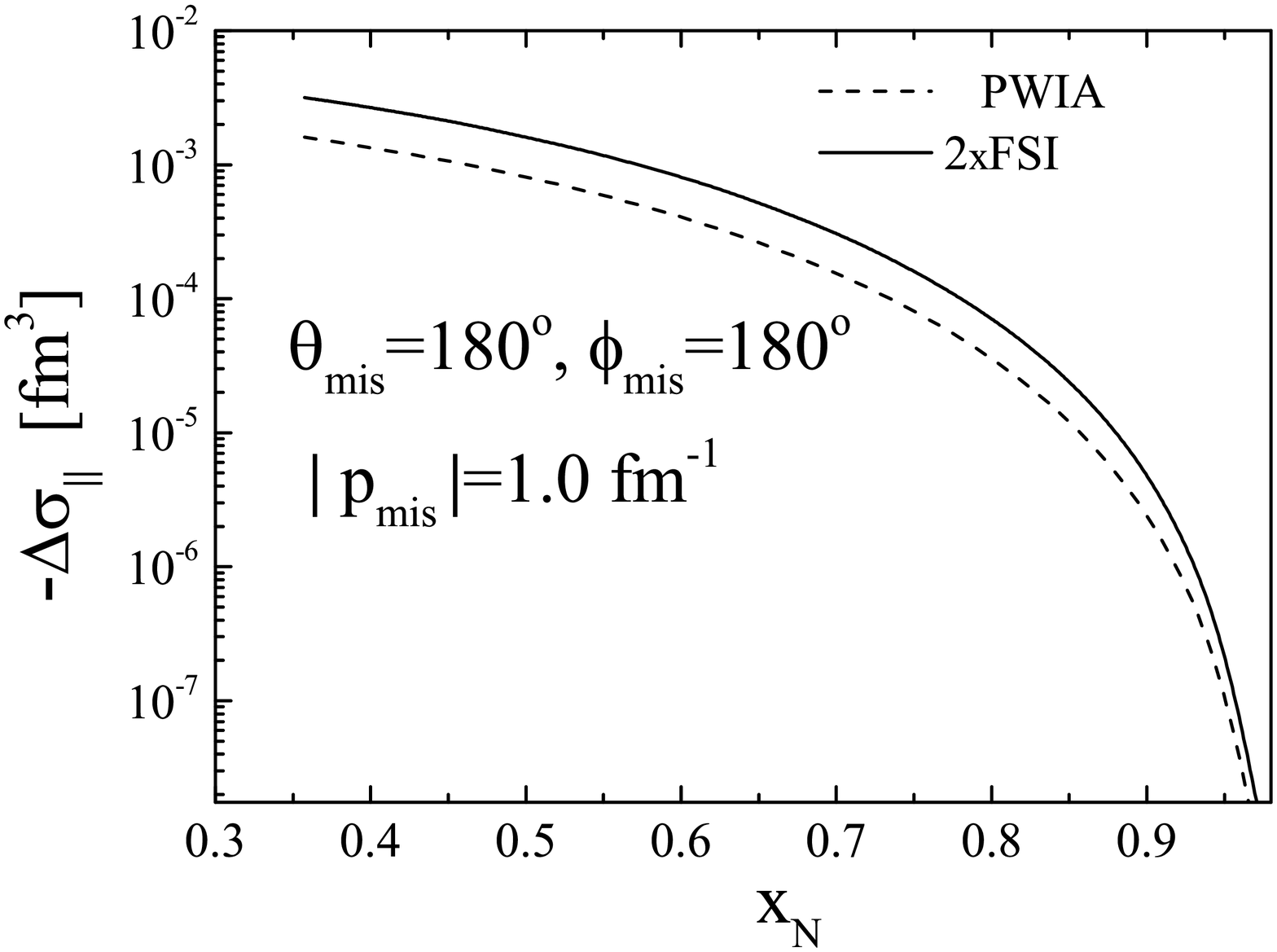}
         \includegraphics[width=0.45\textwidth]{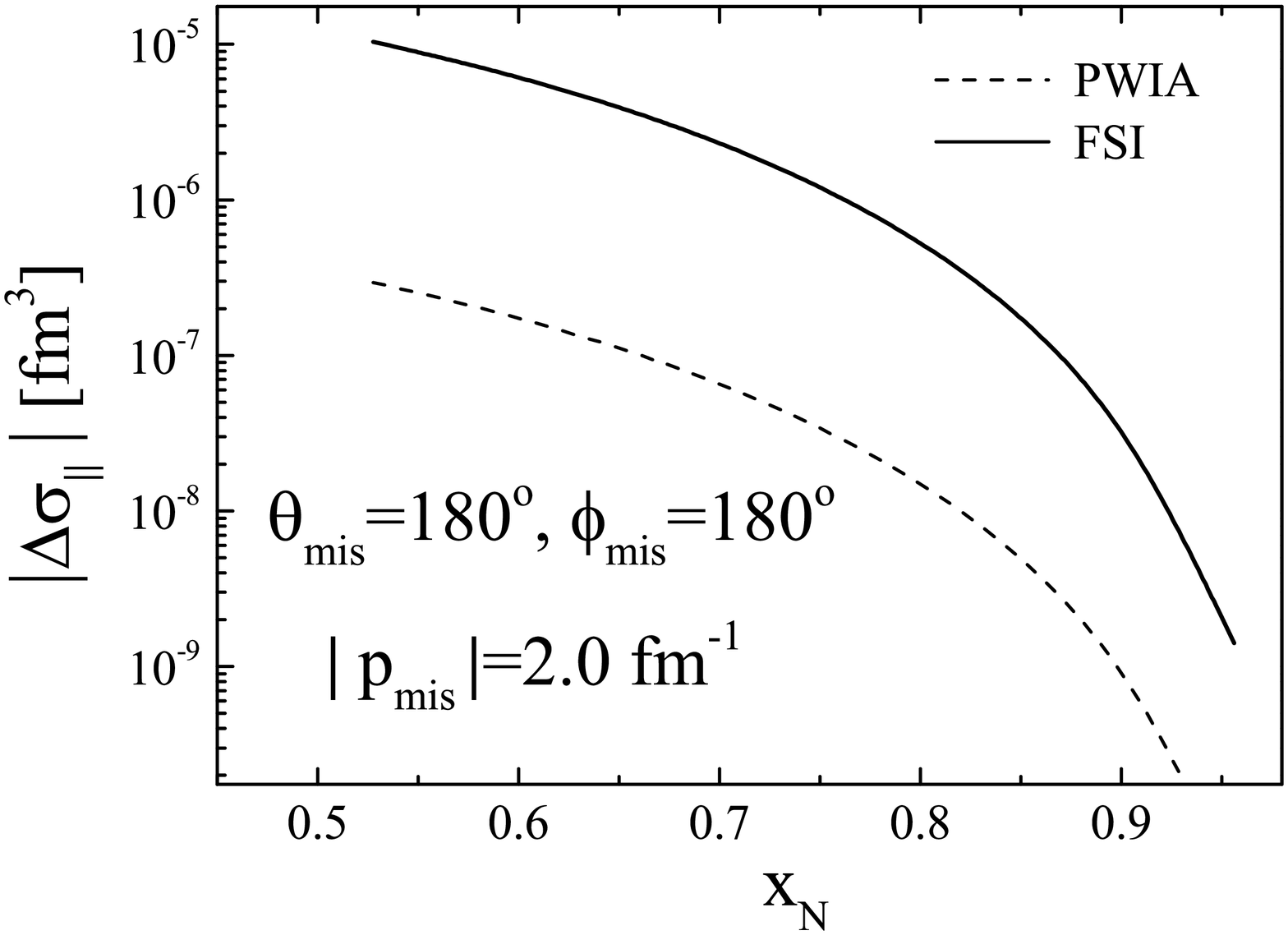}
\caption{
The
 reduced parallel asymmetry, Eq.(\ref{tag}), for the the reaction
  $^3\vec {\rm He}(\vec e,e'~ {^2{\rm  H})}X$  vs  $x_{N}$,  { with a
  longitudinally-polarized target and choosing
     ${\cal E}=12 ~GeV$  and
$Q^2=6~(GeV/c)^2$}. Dashed line: PWIA calculations.
Solid line: results with   FSI effects.
Left panel: calculations  with $|{\bf p}_{mis}|=1 fm^{-1}$, i.e. swhere
 FSI effects are negligible, (cf. Fig.~\ref{figSparPerp}).
Right panel: the same as the left panel, but for  $|{\bf p}_{mis}|=2 fm^{-1}$,
 i.e.
where   FSI effects are maximized and dominate the PWIA calculations
 by two orders of magnitude. { Notice that  in this panel the absolute
 value is
 shown, since
  the sign of the FSI calculation   is negative} (see text).}
                \label{parall}
\end{figure}

In Fig.~\ref{parall}  numerical calculations of the reduced
asymmetry (\ref{tag}) {  for the reaction $^3\vec {\rm He}(\vec e,e' {^2{\rm H})}X$
are presented within the parallel kinematics at fixed values of $|{\bf
p}_{mis}|$, as a function of
$x_N=x_{Bj}/z_N$, } putting ${\cal E}=12 ~GeV$  and
$Q^2=6~(GeV/c)^2$  .
This setting allows one to determine { i)
$p^\mu_N\equiv \{M_{\rm He}-\sqrt{M^2_D +|{\bf p}_{mis}|^2},{\bf p}_{mis}\}$
  ii) $\nu$ through $p_N \cdot q =Q^2/(2 x_N)$ and iii) finally
 $x_{Bj}$ and $z_N$.}
The left panel of Fig.~\ref{parall}  shows the parallel asymmetry at
 $|{\bf p}_{mis} |=1 fm^{-1}$, i.e. in the case where the FSI
corrections are
small (cf Fig. (\ref{figSparPerp}))  and the study of
the proton structure function, $g_{1p}$ is highly
  feasible. { It is very important to note} that the calculations
  with FSI effects
have been arbitrarily multiplied by a factor of two for making it
distinguishable from the PWIA results. It should be pointed out that with
such a kinematical choice, one has
${\cal P}_{||}^{FSI}({\bf p}_{mis})={\cal P}_{||}^{PWIA}({\bf p}_{mis})\approx 1\cdot 10^{-2}$ fm$^3$,
and $0.27< x_{Bj} < 0.73$, so that
$0.73 < z_N < 0.77$.   The right panel illustrates
 the case   at $|{\bf p}_{mis}|=2 fm^{-1}$, where FSI effects are sizable: two order of magnitude larger than the PWIA result.
  It is worth emphasizing that such a huge difference prevents
 a reliable  extraction   of $g_{1p}$,  at high
missing momenta, even if the ${\cal P}^{PWIA}_\parallel$
and  ${\cal P}^{FSI}_\parallel$
remain constant
in the whole range of $x_N$ investigated.
 For  $|{\bf p}_{mis}|=2 fm^{-1}$, the scaling variable $x_{Bj}$ varies in the interval $0.27< x_{Bj} < 0.48$, and
$0.48 < z_N < 0.51$.
{ As a final remark, we point out that, in our calculations,
we have taken into account
the difference between  $\nu$  and $|{\bf q}|$, since in the adopted
 kinematics the Bjorken regime is not fully reached. }
This causes $z_N$ to slightly change with
increasing $x_{Bj}$, at fixed values of $|{\bf p}_{mis}|$ and  $\theta_{mis}$.

In order to tag the { polarized} EMC effect, i.e. if, how,   and to what extent, the nucleon
structure function in the medium differs from the free structure function,
one has to get rid  of  the effects due to the  distorted nucleon { spectral
function} and other nuclear structure effects. One has to consider a quantity
which would depend only upon $g_{1N}(x_{Bj})$.
This can be achieved by considering the ratio of parallel asymmetries, Eq.
(\ref{tag}),
measured at two different values of the Bjorken-scaling variable
$x^{(1)}_{Bj}$ and $x_{Bj}^{(2)}$, leaving unchanged
all the other quantities. {   Therefore,
exploiting the spectator SiDIS by a polarized $^3$He in 2bbu channel,
 information on $g_{1{ N}}(x_{N})$ can be obtained from
the following ratio}
\begin{equation}
R_{||}\left( x_{Bj}^{(1)},x_{Bj}^{(2)}\right)
=\frac{\Delta\sigma_{\|}(x_{Bj}^{(1)},{\bf p}_{mis})}
{\Delta\sigma_{\|}(x_{Bj}^{(2)},{\bf p}_{mis})}
=\frac{g_{1{ N}}(x_{Bj}^{(1)}/z_N^{(2)})}{g_{1{ N}}
(x_{Bj}^{(2)}/z_N^{(2)})}~.
\label{ratio}
\end{equation}
In the Bjorken limit $z_N^{(1)}=z_N^{(2)}$ since
$z_N\approx 1 - |{\bf p}_{mis}|/m_N$ and therefore Eq.~(\ref{ratio}) directly
reflects the $x_N$-dependence of $g_{1 N}$.

Let us now focus on the possibilities of using {  the spectator} SiDIS to get information on
the hadronization mechanism. As already discussed, for such purposes it is
more convenient
to consider the perpendicular kinematics, namely when FSI are dominant.
The kinematics must be chosen in such a way that, by varying the
missing momentum $|{\bf p}_{mis}|$, the argument of the nucleon
structure function $g_{1{ N}}(x_N)$ remains constant.
In this case the reduced asymmetry will depend only on the spectral-function
components $ {\cal P}_{||}^{\frac12  }({\bf p}_{mis})$ and
${\cal P}_{\perp}^{\frac12  }({\bf p}_{mis})$.

\begin{figure}
\includegraphics[width=0.47\textwidth]{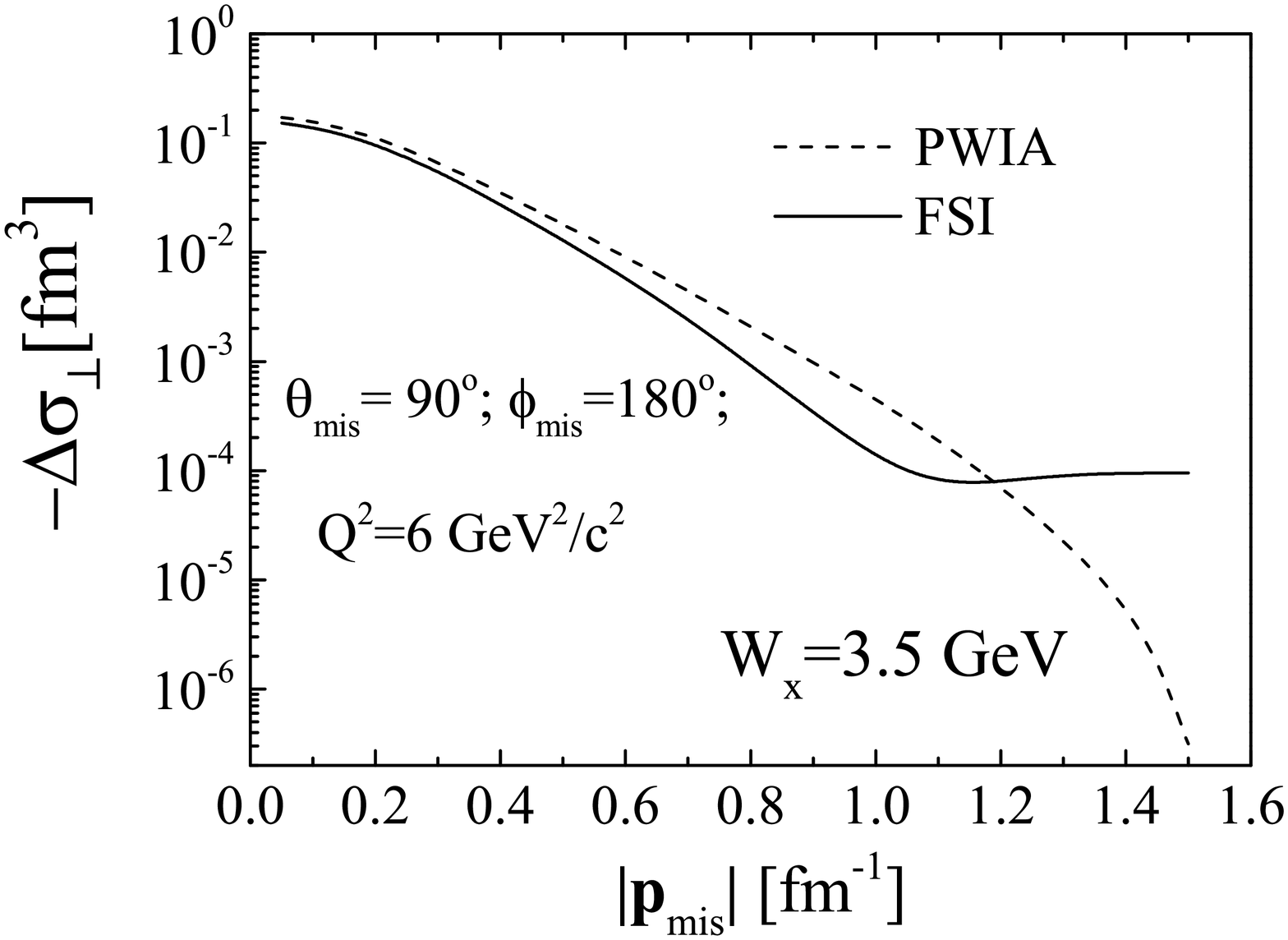}
\includegraphics[width=0.47\textwidth]{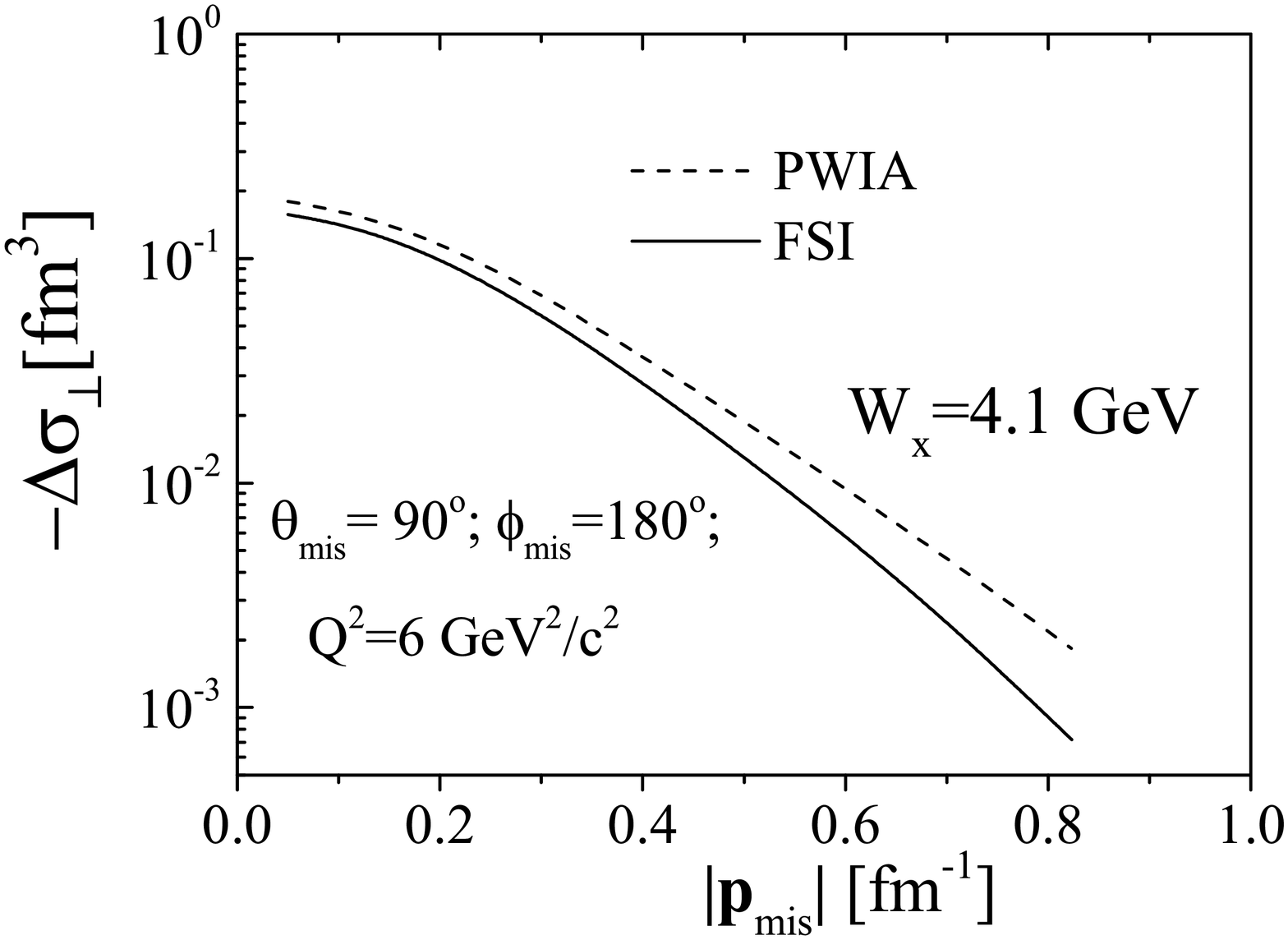}
                \caption{
The reduced { perpendicular} asymmetry, Eq.(\ref{tagPerp}), vs $|{\bf p}_{mis}|$
for the process $^3\vec {\rm He}(\vec e,e'~ {^2{\rm  H})}X$, { with a
  longitudinally-polarized target,  choosing
     ${\cal E}=12 ~GeV$ and 
$Q^2=6~(GeV/c)^2$,  for two values of the invariant mass  $W_X$}. Dashed
line: PWIA results.  Solid lines: results with FSI taken into account.
Left  panel: calculations corresponding to $W_X=3.5 ~GeV$, leading to
$x_N\approx 0.34$. Right panel:
calculations corresponding to $W_X=4.1~ GeV$, that yields $x_N=0.27$.
}
\label{perpendicular}
\end{figure}

In perpendicular kinematics, the term inside the
square brackets in Eq.~(\ref{crosa-21}) becomes { (remind that in perpendicular
kinematics $ \hat {\bf p}_N \cdot {\bm {\cal P}}^{\hat {\bf S}_A} =
{\cal P}_{\perp}^{\frac12  }({\bf p}_{mis})$ and
moreover
$\KE \cdot {\bf p}_N= -|{\bf p}_{mis}|~{\cal E}' \sin\theta_e$)}

\begin{eqnarray}
\left[\phantom{\frac12}\cdots \right]=&&
{\cal P}_{\perp}^{\frac12 \ }({\bf p}_{mis}) \left[ {\cal E}'\sin\theta_e
+\frac{|{\bf p}_{mis}|}{m_N} \left({\cal E} + {\cal E}'\right)
{  -{\cal E}'\sin\theta_e\frac{|{\bf p}_{mis}|^2}{2m_N^2}}
\right]  
+{\cal P}_{||}^{\frac12  }({\bf p}_{mis}) \left[ {\cal E} + {\cal E}'
\cos\theta_e\phantom{\frac12}\right] \nonumber \\&&
\approx
{\cal E} \left[ {\cal P}_{||}^{\frac12  }({\bf p}_{mis})\left({  2-y
-{Q^2 \over 2 {\cal E}^2}}\phantom{\frac12 \!\!\!}\right )
 +{\cal P}_{\perp}^{\frac12  }({\bf p}_{mis})
 \left(2-y\phantom{\frac12 \!\!\!}\right )
\frac{|{\bf p}_{mis}|}{m}\right]\nonumber \\&&
\approx
{\cal E} ~(2-y)~
\left[ {\cal P}_{||}^{\frac12  }({\bf p}_{mis})
 +{\cal P}_{\perp}^{\frac12  }({\bf p}_{mis})
\frac{|{\bf p}_{mis}|}{m_N}\right]~~~~.
\label{perp}
\end{eqnarray}
  where  also the  term proportional to
$ \sin\theta_e~{\cal E}'/{\cal E}$ has been disregarded, since
 in the DIS kinematics it becomes negligibly small.
As in the parallel case, let us define the reduced perpendicular asymmetry

\be
\Delta\sigma_\perp\equiv
\frac{\Delta \sigma^{\hat {\bf S}_A}(\theta_{mis}\sim90^o)}{d\varphi_e dx_{Bj}
 dy d{\bf P}_{D}}
  \left / {\left( \frac{2(2 -y) \, \alpha_{em}^2 \, m_N}{Q^2 E_N} \right) }
\right .=\nonu
\approx g_{1N}\left(\frac{x_{Bj}}{z_N}\right)
{
\left[ {\cal P}_{||}^{\frac12  }({\bf p}_{mis})
 +{\cal P}_{\perp}^{\frac12  }({\bf p}_{mis})\frac{|{\bf p}_{mis}|}{m_N}\right]~~~~.
}
\label{tagPerp}
\ee

In Fig.~\ref{perpendicular} the reduced  asymmetry for  a spectator
SiDIS by a polarized $^3$He target, with a detected deuteron, is presented
as a function of the missing momentum in the perpendicular kinematics.
In this case, one arranges the  other kinematical variables in such a way that
$x_N$ remains constant in the whole range of the explored $|{\bf p}_{mis}|$.
In particular, once $|{\bf p}_{mis}|$ is given, then the four-momentum $p^\mu_N$ is
determined  (see  above), and after fixing both $Q^2$ and the  invariant
mass of the debris, $W_X^2=(p_N+q)^2$, one can straightforwardly determine
 $x_N= Q^2/(W^2_X-p^2_N+Q^2)$, and in turn $y$. Since the considered
 $|{\bf p}_{mis}|$ is at
 most $1.5 fm^{-1}$, one can immediately realize that $p^2_N \sim m^2_N$, and
 therefore $x_N$ is almost constant in the whole range of $|{\bf p}_{mis}|$.
  From  Fig.~\ref{perpendicular},  it is seen
that FSI effects increase with $|{\bf p}_{mis}|$ and, for
{$|{\bf p}_{mis}|> 0.6\ fm^{-1}$}, become dominant. Therefore,
 by measuring the perpendicular
asymmetry and comparing with the PWIA calculations one can get information
on the magnitude of FSI. Since FSI are determined by the effective
debris-nucleon cross section, this should provide information on
$\sigma_{eff}$, and  in turn
 on the hadronization mechanism. Notice that the difference between $x_{Bj}$
and $x_N$ is due to the off-mass-shellness of the hit nucleon,
i.e. $p^2_N\ne m^2_N$, but
at the considered values of
$|{\bf p}_{mis}|$, this difference is negligibly small. { Moreover, in the whole range
of $|{\bf p}_{mis}|$ shown in Fig. ~\ref{perpendicular}, the value of $x_N$
remains almost constant.}
 In the left panel,  $x_N\approx 0.34$ and
the{  proton } structure function amounts to
  $g_{1{ p}}(x_N)\approx 0.174 $.

In the right panel, the value of the invariant mass { corresponds to an
energy transfer close to the experimental kinematical limit ($\nu\approx 11.8\  GeV$) and
it is reached when
 electrons scatter}
in the backward hemisphere. For this reason  $|{\bf p}_{mis}|$
is kinematically
constrained, with a cut $< 1 fm^{-1}$.  Moreover one has  $x_N=0.273$  and
$g_{1{ p}}(x_N)\approx 0.218 $.
 { As in the left panel, both PWIA and FSI nuclear-structure contributions
 are negative.}



\subsection{Discussion}
 The previous  calculations,  regardless of the direction of
the polarization vector ${\bf S}_A$, show that
all partial  components of the spin-dependent spectral function,
Eqs. (\ref{cyclic}), (\ref{partial}) and
(\ref{kkk}), are different from zero, once FSI effects are taken into account. However, as expected,
the components along ${\bf S}_A$ are much larger than the ones
in the perpendicular direction (cf Figs. (\ref{figSparPerp}) and
(\ref{SpinPerp})). When the asymmetries, Eq. (\ref{crosa-21}), are considered,
one can see that the
perpendicular component of the spectral function can appear only in
 scalar-product combination with
 ${\bf p}_N$. This leads to a  further   reduction, by a factor
 $|{\bf p}_N|/m_N$. Therefore,  the
perpendicular component of the spectral function  can be safely neglected
in the spectator SiDIS by a proper
choice of kinematics. But, in general, the contribution
of components perpendicular to the nuclear polarization vector cannot be
"a priory" disregarded.
In particular, if one  considers a   standard SiDIS reaction, i.e.
{ with a fast hadron detected, and a transversely-polarized $^3$He
target, i.e. $\hat {\bf S}_A \perp \hat {\bf k}_e\sim \hat {\bf q}$, one should pay attention to FSI effects, since
the contribution of the perpendicular components to the asymmetries
 can be, in principle, sizable. The above mentioned SiDIS process is relevant
 for extracting the neutron transversity, but also Collins and Sivers functions,
 and therefore it has to be carefully
 investigated. This  analysis will be presented elsewhere \cite{tobe}, but from
 the present study one can extract  some general considerations and develop
 some
 expectations.
 In the transversely-polarization setting of the target, the nuclear tensor
 contains contributions from both diagonal
 and non diagonal terms, viz (cf Eq. (\ref{roty}))
 \be
 W_{\mu\nu}^{s.i.}(S_{A\perp}, Q^2,P_h)={1 \over 2}\ W_{\mu\nu}^{\frac12\frac12}+\
{1 \over 2} W_{\mu\nu}^{-\frac12-\frac12}+ \ \left[\frac12\left(
W_{\mu\nu}^{\frac12-\frac12}+W_{\mu\nu}^{-\frac12\frac12}\right)\right]~.
\label{rotyt}
 \ee
 This entails that one has to
 calculate the whole polarization contribution to the spin-dependent spectral
 function,  as given  in Eqs. (\ref{partial}) and (\ref{nond}), in order to get
 information on FSI effects in standard SiDIS. Moreover,
in standard SiDIS, the $(A-1)$ nucleus is not detected and one needs to integrate
over all the values of the missing momentum, see Eq. (\ref{crosa-2}).  Indeed, since
in the present
paper we have evaluated the diagonal ${\bm {\cal P}}^{\cal MM}_{(FSI)}$,
one can infer that there should be kinematical ranges where FSI could modify the
PWIA cross sections, but not necessarily the asymmetries. In particular,
Fig. (\ref{figSparPerp}), where the diagonal $| {\cal P}^{\frac12}_\parallel({\bf
p}_{mis})|$ is shown for the parallel and perpendicular kinematics,
indicates the kinematical region where one should expect large FSI effects,
 (remind that the diagonal ${\cal P}^{\frac12}_\perp({\bf
p}_{mis})$ is always quite small)}.
Here we notice that, in principle, FSI effects on the cross section could be properly
{minimized } if one measured a more exclusive SiDIS process,
{ where both  the $(A-1)$ spectator and the fast hadron are detected,}
i.e by considering the reaction $l+\vec A= l'+h+(A-1) +X$ with a fast hadron
$h$ and a slow $(A-1)$ nucleus.

\section{Conclusions and Perspectives}
{ In summary, we introduced a  novel nuclear distribution function
 for investigating the semi-inclusive deep-inelastic scattering of polarized leptons
by a polarized $A=3$ nucleus. Such a distribution function,
the distorted spin-dependent spectral function, has been evaluated
for an $A=3$ nucleus, by using the three-body wave function \cite{pisa} corresponding
to the realistic AV18 NN interaction \cite{av18}.
 Our calculations
should make more reliable the extraction of information on a bound nucleon,
with a particular
aim of providing a refined treatment of a polarized $^3$He target, that, as well-known, represents an effective
neutron target, and plays a fundamental role for obtaining observables
for the neutron like  i) transversity
and ii) Collins and Sivers functions, i.e. the main physical motivations of
forthcoming experiments with a transversely-polarized
$^3$He \cite{He3exp}.

As a first step, we gave the formal expression of the nuclear tensor in terms of the nuclear
overlaps, i.e. the transition amplitude between the target wave function and a state given by the Cartesian product
of a nucleon plane-wave and the fully-interacting state composed by  the remaining $(A-1)$ nucleons.
Within such a formalism the
nuclear tensor can be factorized in a nucleonic tensor and the proper combination of the above mentioned overlaps,
that in turn yield the well-known  PWIA spin-dependent spectral function. The FSI effects have been introduced
through a
generalized eikonal approximation, successfully applied to the description of  unpolarized spectator
SiDIS processes \cite{ourlast}. The spectator SiDIS reactions,  where a slowly-recoiling $(A-1)$ systems is
detected in coincidence with the scattered lepton, differently from the
standard SiDIS, where a fast hadron,
produced after the virtual-photon absorption by a quark inside the bound
nucleon is
detected, have been chosen as a test-ground for our approach.
In particular, we have applied our formalism to the reaction $^3\vec {\rm He}(\vec e,e' ~{^2{\rm  H})}X$, with a
longitudinally-polarized target. The investigation of the
spin-dependent spectral function  needed for describing the previous spectator SiDIS
has allowed to single out a kinematical region where the FSI can be minimized, and therefore one can safely extract
the
polarized structure function of a bound proton, $g_{1p}$, and a region where 
the FSI has  a maximal effect. There,
 one can
address the issue of the hadronization in a nuclear medium. The analysis of 
this spectator SiDIS,
 though less complicate and
less numerically-demanding than the standard one, provide valuable information on a nucleon bound in a nucleus, shedding some light on
the polarized EMC effect, and moreover allows us to calibrate our approach
for the next application to the
standard SiDIS by a transversely-polarized $^3\vec {\rm He}$.
In this case we have to extended the calculation of the distorted spin-dependent spectral function to the
off-diagonal terms, i.e. to the combination of the nuclear overlaps corresponding to target states with different
spin projections, and then to integrate over a suitable (given by the experimental  kinematics) range
of missing momentum
and energies of the $(A-1)$ system. Seemingly, this task  appears highly non  trivial from the numerical point
of view,
while from the theoretical side  the whole formalism can be derived following Eq. (\ref{prop}) and generalizing
Eq. (\ref{lp}).
Coming back to the spectator SiDIS, the approach presented in this paper, and the numerical results for the
spin-dependent spectral function (cf Figs. \ref{figSparPerp},
\ref{spinDeOtugla} and  \ref{SpinPerp}), can be
applied to a polarized Tritium target,
since for the time being we have disregarded the Coulomb effects.
Through the reaction  $^3\vec {\rm H}(\vec e,e'~ {^2{\rm  H})}X$ one could address the issue of the structure
function of a bound neutron, $g_{1n}$ and, analogously to the $^3$He case, the hadronization process involving
a quark inside a
bound neutron. In spite of the apparent difficulties to get a
a polarized $^3$H target,  one could conceive its construction as not
impossible  in a not too distant future, since in the last decade
 an  impressive amount of R\&D achievements has made  available an
 unpolarized Tritium target for planned and approved
 experiments \cite{marathon}.}

\section*{Acknowledgments}
 This work was   supported
  by the Research Infrastructure Integrating Activity Study of Strongly Interacting Matter
  (acronym HadronPhysic3, Grant Agreement n. 283286) under the Seventh Framework
Programme of the European Community and partially done under
 U.S. DOE Contract No. DE-AC05-06OR23177. Calculations were in  part
performed on Caspur facilities under the Standard HPC 2012,
STD12-141 grant "SRCnuc".   We thank M. Alvioli for relevant support in running MPI codes.~
 L.P.K. thanks the financial support from  the JLab theory group and
the program "Rientro dei Cervelli" of the Italian Ministry of University and Research. Moreover, L.P.K.
acknowledges the warm hospitality of the JLab theory group
and Patrizia Rossi for  her very friendly support.

\end{document}